\documentclass[manuscript]{acmart}
\usepackage{booktabs}
\usepackage{threeparttable}
\usepackage{multirow}

\setcopyright{acmcopyright}
\copyrightyear{2024}
\acmYear{2024}
\setcopyright{rightsretained}
\acmConference[CHI '24]{Proceedings of the CHI Conference on Human Factors in Computing Systems}{May 11--16, 2024}{Honolulu, HI, USA}
\acmBooktitle{Proceedings of the CHI Conference on Human Factors in Computing Systems (CHI '24), May 11--16, 2024, Honolulu, HI, USA}
\acmDOI{10.1145/3613904.3642038}
\acmISBN{979-8-4007-0330-0/24/05}

\begin{document}
\title[RobotVoice]{Giving Robots a Voice: Human-in-the-Loop Voice Creation and open-ended Labeling}
\author{Pol van Rijn}
\email{pol.van-rijn@ae.mpg.de}
\orcid{0000-0002-4044-9123}
\affiliation{%
  \institution{Max Planck Institute for Empirical Aesthetics}
  \city{Frankfurt}
  \country{Germany}
}

\author{Silvan Mertes}
\email{silvan.mertes@uni-a.de}
\orcid{0000-0001-5230-5218}
\affiliation{%
  \institution{Chair for Human-Centered Artificial Intelligence, University of Augsburg}
  \city{Augsburg}
  \country{Germany}
}

\author{Kathrin Janowski}
\email{kathrin.janowski@uni-a.de}
\orcid{0000-0001-5985-4973}
\affiliation{%
  \institution{Chair for Human-Centered Artificial Intelligence, University of Augsburg}
  \city{Augsburg}
  \country{Germany}
}

\author{Katharina Weitz}
\email{katharina.weitz@uni-a.de}
\orcid{0000-0003-1001-2278}
\affiliation{%
  \institution{Chair for Human-Centered Artificial Intelligence, University of Augsburg}
  \city{Augsburg}
  \country{Germany}
}

\author{Nori Jacoby}
\email{nori.jacoby@ae.mpg.de}
\orcid{0000-0003-2868-0100}
\authornote{Both authors contributed equally to this research.}
\affiliation{%
  \institution{Max Planck Institute for Empirical Aesthetics}
  \city{Frankfurt}
  \country{Germany}
}

\author{Elisabeth André}
\authornotemark[1]
\email{elisabeth.andre@uni-a.de}
\orcid{0000-0002-2367-162X}
\affiliation{%
  \institution{Chair for Human-Centered Artificial Intelligence, University of Augsburg}
  \city{Augsburg}
  \country{Germany}
}

\renewcommand{\shortauthors}{Van Rijn, et al.}

\begin{abstract}
Speech is a natural interface for humans to interact with robots. Yet, aligning a robot's voice to its appearance is challenging due to the rich vocabulary of both modalities. Previous research has explored a few labels to describe robots and tested them on a limited number of robots and existing voices. Here, we develop a robot-voice creation tool followed by large-scale behavioral human experiments (N=2,505). First, participants collectively tune robotic voices to match 175 robot images using an adaptive human-in-the-loop pipeline. Then, participants describe their impression of the robot or their matched voice using another human-in-the-loop paradigm for open-ended labeling. The elicited taxonomy is then used to rate robot attributes and to predict the best voice for an unseen robot. We offer a web interface to aid engineers in customizing robot voices, demonstrating the synergy between cognitive science and machine learning for engineering tools.
\end{abstract}



\begin{teaserfigure}
  \includegraphics{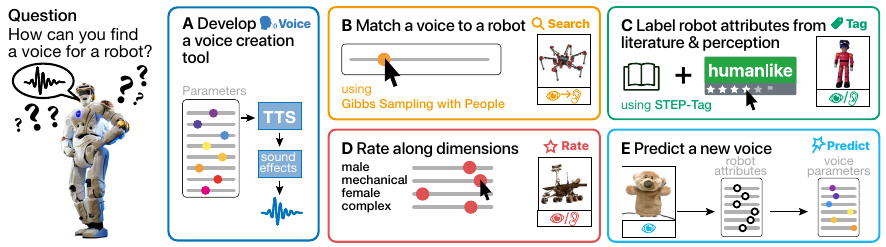}
  \caption{How can you find an appropriate voice for a robot? We propose a five-step approach: \textbf{A} Develop a voice creation tool for robots. \textbf{B} Participants iteratively change the voice of the robot using this tool to find a voice that fits well to the robot. \textbf{C} Identify attributes relevant to the perception of robots from previous literature and a taxonomy elicitation procedure. \textbf{D} A separate set of participants rates all images and their matched voices along those perceptual attributes. \textbf{E} Predict well-matched voices for unseen robots. For this and all future figures, the copyright holders of the robot images are defined in Tables \ref{stab:robot_selection1}--\ref{stab:robot_selection2}.}
  \Description{How can you find an appropriate voice for a robot? We propose a five-step approach: A: Develop a voice creation tool for robots. The subfigure shows a schematic interface of sliders and how they are fed into a text-to-speech model and into a sound effects module that produces the voice. B: Participants iteratively change the voice of the robot using this tool to find a voice that fits well to the robot. The subfigure shows a window with a slider in it. By moving the slider one voice dimension is manipulated using Gibbs Sampling with People. From this process done on images, we obtain matched voices. C: Identify attributes relevant to the perception of robots from previous literature and a taxonomy elicitation procedure. The subfigure shows the interface used for STEP-Tag and an open book signifying existing literature. STEP-Tag is run on the image and voice modality separately. D: A separate set of participants rates all images and their matched voices along those perceptual attributes. The subfigure shows multiple sliders at once to the participant to rate perceptual dimensions. The rating experiment is run on the image and voice modality separately. E: Predict well-matched voices for unseen robots. The subfigure shows the schematics of obtaining a predicted voice: From an image, we retrieve the perceptual ratings of the robot's attributes. From the perceptual ratings, we predict voice parameters.}
  \label{fig:intro}
\end{teaserfigure}

\maketitle

\section{Introduction}

Robots are used in a wide range of scenarios and they vary in purpose and appearance \cite{mutlu:et:al:2006}. The voice is an intuitive medium for humans to interact with robots, conveying not only spoken content but also intentions \cite{ponsot2018intention}, personality \cite{scherer1978personality}, conversational goals \cite{kohler2011goals}, and emotions \cite{banse1996emotion}. However, a discrepancy between what we see (the robot's appearance) and what we hear (its voice) can strongly hinder robots’ usability. Previous research has stressed the importance of users' affective responses to robots in fulfilling their functions \cite{breazeal:2004, Janowski2022}. However, a mismatched voice can result in a variety of aversive reactions, such as unsettling, eerie, uncanny, and repulsive responses \cite{Mitchell2011, Meah2014, Tinwell2015,mori2012uncanny, schwind2018uncanny}. The intensity of this dissonance can be influenced by factors like the user's age or the robot's realism \cite{esposito2019dependability, mertes2021potential}.

Given the broad spectrum of robot designs, the range of possible voices given to robots must be similarly diverse. For example, functional robots need to be highly intelligible (e.g., for tasks like navigation), whereas popular media robots, like Pixar's \textit{WALL-E}, are designed to sound less clear but more expressive. However, existing robot voices are limited in their diversity \cite{Cambre:Kulkarni:2019} and research is often limited in terms of the number of explored voice dimensions \cite{james2020empathetic, torre2020if, torre2018trust}, the number of  robots studied \cite {
alonso2019online, haering:et:al:2014,  james2020empathetic, torre2020if, torre2018trust}, or in terms of the robots' diversity \cite{Perugia2022}. So, how can we synthesize a robot voice, accounting for the broad spectrum of possibilities?

The advancement of Text-To-Speech (TTS) systems has made it possible to synthesize realistic human voices \cite{tan2021review}. However, desirable robot voices may significantly differ from human voices. This work extends a state-of-the-art TTS model \cite{kim2021vits} (Figure \ref{fig:intro}A) to cover a wide range of robot voices: from highly synthetic or distorted voices to natural and individualized voices that sound similar to human speech. How can we efficiently search the expansive space of all voices to find the one that best matches a newly crafted robot?
For this problem, we use an adaptive, human-in-the-loop sampling paradigm (Gibbs Sampling with People, or GSP;  \cite{harrison2020gsp})  to iteratively find a voice that fits a robot (Figure \ref{fig:intro}B). 
803 participants engaged in the task of matching a voice to 175 images of commonly used robots that span a wide variety of appearances and contexts. A separate group of human raters (\textit{N} = 142) confirmed that the created voice improves over iterations and plateaus in the last iterations.

We then performed a literature review and identified attributes that characterize robots and their voices. We compared this list to labels elicited directly from participants viewing images of robots (\textit{N} = 73) or listening to their matched voices (\textit{N} = 59). To do so, we use a recently developed adaptive human-in-the-loop labeling paradigm \cite{marjieh_words_2022} that does not rely on a pre-existing taxonomy (see Figure \ref{fig:intro}C). We found that terms emerging from this process mostly overlapped with those proposed in the literature. 
We then compiled a new list of 40 labels that frequently appear both in the literature and in our labeling pipeline and recruited new groups of participants to rate the voices (\textit{N} = 245) and images (\textit{N} = 298) along these dimensions (see Figure \ref{fig:intro}D).
Finally, we show that the perceptual rating of the image predicts a suitable voice for a robot (see Figure \ref{fig:intro}E). 
A separate group of raters confirmed that the predicted voice is similarly good as the matched voice (\textit{N} = 94). We conducted two separate experiments on a new set of 175 robot images from the ABOT dataset \cite{Phillips2018} (\textit{N} = 249) and on randomly generated voices (\textit{N} = 189) to ensure the reliability of our results. We observed that the relationship between labels in these new datasets was similar to the original one. Furthermore, using the ratings of the new robots we propose a matched voice optimized from images in the first set. We show the predicted voice is as good as the original match, confirming the robustness of our findings. We have made the developed voice creation tool publicly available as a Python package\footnote{\url{https://robotvoice.s3.amazonaws.com/code.zip}} to enable our validated voice configurations to be directly used in real-world applications. Finally, we provide an online robot voice prediction tool, which can be used to identify possible voices for new robots\footnote{\url{https://robotvoice.s3.amazonaws.com/predict.html}}.

The contributions of this work can be summarized as:
\begin{itemize}
    \item We provide a voice creation tool that covers a wide range of robotic voices using both state-of-the-art TTS and classical signal processing (Figure \ref{fig:intro}A).
    \item We present a human-in-the-loop approach for creating a synthetic voice for a particular robot (Figure \ref{fig:intro}B).
    \item We use the taxonomy elicitation process to identify labels that are relevant for the perception of robots, both in audition and vision, and compare them with attributes from the literature (Figure \ref{fig:intro}C).
    \item We create a densely annotated dataset of the attributes of 175 robots (Figure \ref{fig:intro}D). 
    \item We show how our tool predicts suitable voices for new robots based on those perceptual dimensions (Figure \ref{fig:intro}E).
    \item In order to demonstrate that our results are robust regardless of the initial set of robots, we rerun the annotation and prediction steps with a different set of 175 robots.
    \item We make the resulting TTS voices publicly available as an easy-to-use software package.
\end{itemize}

\section{Related Work}
Here we first review related research exploring the correlation between a robot's appearance and voice dimensions, underlining the significance of aligning a robot's voice with its perceived attributes. We then propose to apply two recently developed approaches to human-in-the-loop alignment to robot voice alignment: 1) human-in-the-loop sampling \cite{harrison2020gsp}, which is the foundation of our method for collectively generating voice samples that match a robot's appearance, and 2) human-in-the-loop labeling \cite{marjieh_words_2022}, which we use to capture people's auditory or visual perceptions of robots.

\subsection{Robots and Speech}
Existing TTS models have frequently been used as voices for robots \cite{roehling2006towards} and their quality has greatly improved in the last decade \cite{tan2021review, communication2023seamlessm4t}, enabling them to produce speech that is nearly indistinguishable from human recordings \cite{kim2021vits}. humanlike voices are typically preferred over synthetic ones \cite{kuhne2020human}, which makes state-of-the-art TTS models an excellent voice creation tool. 
A recent switch from recurrent to non-autoregressive models \cite{kim2020glow, valle2020flowtron, siuzdak2022WavThruVec} brought about major improvements in latency, allowing robots to produce voices faster than real-time. Modern TTS models have great factorization abilities \cite{wang2018GST, valle2020mellotron, lancucki2021fastpitch}, allowing users to independently change text, prosody, and speaker identity (i.e., \textit{what} and \textit{how} something is being said by \textit{whom}). Harnessing such rich latent features \cite{schiller2021} not only facilitates the crafting of new voice personae \cite{jia2018transfer, stanton2021speaker}, but also ensures that these synthesized voices encapsulate the nuances and diversity inherent to human speech.

A substantial body of extant work emphasizes the importance of synchronizing the robot's voice with its appearance \cite{alonso2019online, alonso2020four}.
Simply adopting a TTS model that delivers humanlike speech might be incongruous for a robot that has a distinctly non-human appearance. For example, imagining \emph{R2D2} from Star Wars speaking with a plain natural voice would be odd and likely uncanny \cite{schwind2018uncanny, mori2012uncanny}. 

So what makes a voice appropriate for a robot? Existing studies have investigated this question by looking into the correlation between appearance and voice along certain dimensions (e.g., gender or naturalness).  For example, McGinn et al. \cite{mcginn2019can} developed a voice association task (i.e., matching a picture of a robot to a voice) showing that gender and naturalness strongly affect the visual appearance people associate with a robot. 
Other studies have investigated the opposite relationship: How the voice influences the mental model that people have of a robot. For example, Powers et al. \cite{powers2006advisor} showed that participants associate a male voice with a more knowledgeable person. This research also highlights the risks of reinforcing existing social biases when matching specific vocal characteristics, like a deep voice, with particular personality traits, such as being knowledgeable.
In addition to aligning the voice and appearance of a robot, its behavior must also be synchronized. Torre et al. show that while trust partly depends on the voice \cite{torre2020if}, the consistency of voice and behavior is more important \cite{torre2018trust}. 
This is in line with previous research showing that people prefer serious-sounding robots in work-related contexts \cite{goetz2003matching} and empathetic voices for healthcare robots \cite{james2020empathetic}.

As this literature review emphasizes, aligning robot voices with their appearances is a task of crucial importance. Here, we propose a method that can handle both highly synthetic and natural-sounding robot voices. We also develop a framework to solve the alignment problem by optimizing the voice of a specific robot based on its appearance. Finally, we enrich the aforementioned literature by providing attributes of robots that are relevant to the human perception of voices and images.

\subsection{Human-in-the-Loop Sampling}
Given the wide variety of possible vocal characteristics, tuning a robot's voice is a considerable challenge. While thus far this task has been left to specialists \cite{kondo2019human, mosqueira2023human}, an alternative approach is the human-in-the-loop method.  
Human-in-the-loop methods efficiently integrate human decision-making with computer algorithms, so that a complex computation such as sampling or optimization can be collectively performed by humans and computers \cite{sanborn2010, harrison2020gsp}. Human-in-the-loop techniques have been proposed in the context of mapping internal representation in visual memory \cite{langlois2021}, 3D pose perception \cite{langlois:et:al.2019}, color perception \cite{xu2013}, musical rhythm and melody \cite{jacoby2017, angladatort2022}. More recently, various human-in-the-loop techniques have been developed in the context of speech, including human-in-the-loop evolutionary algorithms to maximize the emotional content in sound \cite{ritschel2019personalized, vanrijn2022gap} and a GUI-based tool allowing users to build a custom TTS voice \cite{kondo2019human}. These approaches are more efficient than elicitation methods that do not use optimization algorithms, such as reverse correlation \cite{dotsch2011, mangini2004, harrison2020gsp}.

\begin{figure}
  \includegraphics{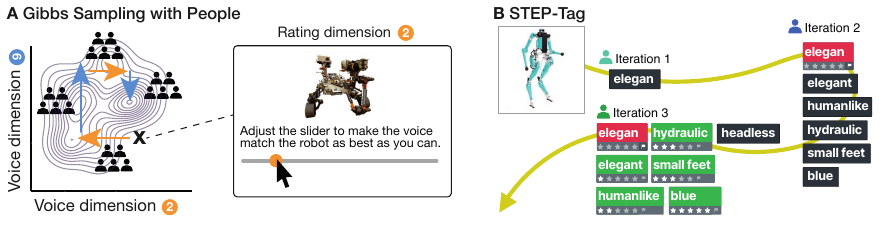}
  \caption{Human-in-the-loop paradigms. \textbf{A} Gibbs Sampling with People. Participants change the slider, modifying only one dimension at a time. By cycling over the dimensions, participants explore dense regions in the feature space that are associated with a given robot. \textbf{B} STEP-Tag. Through the labeling process, participants simultaneously create new tags and review the tags provided by others. Over many iterations, meaningful and rich semantic labels are efficiently collected for each robot image.}
  \Description{A: Schematics of Gibbs Sampling with People. The image shows how one user moves a slider to change one voice dimension at a time. Each slider is presented to five people. The median response propagates to the next iteration. Over the course of iterations, participants explore dense regions in the feature space that are associated with a given robot. B: Schematics of STEP-Tag. Schematics shows an image of a robot and shows the tags assigned to it by different participants.}
  \label{fig:hitl_paradigms}
\end{figure}

A particularly efficient method for optimizing a stimulus for a desired subjective property is Gibbs Sampling with People (GSP) \cite{harrison2020gsp}. In this paradigm, participants are introduced to a stimulus space and use a slider interface to change one dimension of the stimulus space at a time. Importantly, the result from one iteration becomes the input for another iteration, where the same participant or a different participant now manipulates another dimension of the space (Figure \ref{fig:hitl_paradigms}A). For instance, when provided with a robot image, participants might adjust specific voice synthesis parameters sequentially to best align with the given image. Harrison et al. \cite{harrison2020gsp} demonstrated that, under experimentally verifiable conditions, this iterative method converges to samples of high subjective quality - that is, it identifies a voice that perceptually aligns with the image. GSP was previously used in the domain of emotional speech,  \cite{vanrijn21_gsp_gst,harrison2020gsp} or associating a particular voice to a face \cite{vanrijn2022voiceme} in high-dimensional latent spaces in TTS models. To increase the speed of convergence, one can show the same slider to multiple participants and aggregate their responses (e.g., mean or median). In our experiment, we used decisions from 5 participants for every iteration.

\subsection{Human-in-the-Loop Labeling}

One of the best-known models for characterizing human personality is the Big Five personality taxonomy \cite{mccrae_introduction_1992}. Following the Computer as Social Actors paradigm \cite{nass:et:al:1994}, researchers have applied social categories, such as gender, age, and personality, to socially interactive computer agents \cite{kuchenbrandt:et:al:2014, Janowski2022}. Furthermore, specific dimensions have been developed for speech-based conversational agents \cite{voelkel_developing_2020} and for robots \cite{bartneck_godspeed_2009}. Despite these efforts, it remains unclear which dimensions people utilize in their perception of robots and whether these dimensions align with a range of established theories.

A recently developed adaptive tag mining pipeline called Sequential Transmission Evaluation Pipeline (STEP-Tag), where participants adaptively annotate a set of target stimuli, both by providing new descriptive tags for the stimulus and by simultaneously reviewing the tags made by previous participants \cite{marjieh_words_2022}. When the pipeline is applied to images of robots (see Figure \ref{fig:hitl_paradigms}B), participants view the image of a robot, provide tags describing their impression of the robot, and rate the relevance of tags that were created by other participants (5 Likert-scale). Participants also have the possibility of flagging tags they deem inappropriate. Tags are removed if they are flagged twice (but can potentially reappear if a future participant adds them again). As the process unfolds over many iterations, meaningful tags emerge that describe the stimulus well and are validated by multiple participants, thus enabling a theory-free elicitation of tags describing the stimulus. It has previously been demonstrated that this method is effective in eliciting open-ended taxonomies without pre-specification, predicting downstream tasks (such as perceptual and semantic similarity), as well as predicting similarity in the representation of humans and deep learning models \cite{marjieh_words_2022}.

Unlike conventional methods in the literature, STEP-Tag eliminates the need for manual post-processing tasks like merging synonyms, thereby reducing the potential for subjectivity. However, this also means that the provided labels by the user can reflect stereotypes and reveal biases present in the data (e.g., images of cleaning robots often look feminine) and in the participants (e.g., images of masculine-looking robots are perceived as intelligent). Using STEP-Tag can help minimize prejudice in human-robot interactions by characterizing and identifying them, thus enabling engineers to create less biased systems.

\section{Methods}
\subsection{Images of Robots}
As robots vary greatly in their appearance, our goal was to collect a variety of images that capture this variation. To simplify the complexity of possible presentation methods (such as images, videos, and 3D designs), we decided to focus on static images. We used an existing dataset (IEEE Robots) downloaded all robots from \url{https://robots.ieee.org/robots} (April 2022), removing robots without a frontal view and discarded devices such as exoskeletons or telepresence interfaces, which integrate a human user. For each robot, we selected the best image, ideally showing the entire robot in isolation. The selected images span diverse types of robots with 14 different categories, ranging from industrial to consumer robots and humanoids to drones (see Supplementary Materials \ref{sup_image_selection} for the distribution of these categories).

This list of 160 IEEE robots was extended with 15 images that were collected from other sources, such as promotional pictures from manufacturer websites or photographs taken by ourselves. To avoid contextual cues, we removed the shades and backgrounds for all robots and replaced them with a solid white background. In total, we gathered 175 images of robots across many application domains (see Table \ref{stab:robot_selection1}--\ref{stab:robot_selection2}). This selection of 175 robots is notably larger than datasets in relevant previous literature (maximally eight different robots, see Supplementary Materials \ref{sup_number_robots})

\subsection{Voice Manipulation and Effects}
\label{methods_voice}
\begin{figure}[ht!]
  \includegraphics{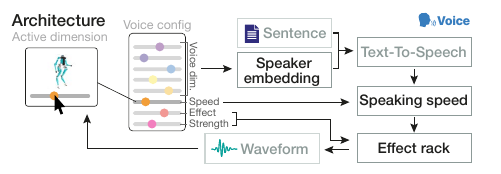}
  \caption{Architecture. The voice of the robot is controlled via eight sliders. The first five sliders control the voice of the TTS model using the first five PCA dimensions on the speaker embeddings. The sixth slider controls the speed of the speech. The seventh slider selects one of the eight effects. The last slider determines the strength of the effect. When moving the slider, the voice configuration updates one parameter in the voice configuration (here: speed). This triggers the synthesis pipeline and the resulting audio is played back to the user.}
  \Description{The figure depicts more detailed schematics of the architecture to produce the robot voice. There are eight sliders that control the voice. The first five sliders control the voice of the text-to-speech model using the first five PCA dimensions on the speaker embeddings. The speaker embedding is fed into the text-to-speech with a given sentence which produces an audio clip. The sixth slider additionally controls the speed of the speech. The seventh slider selects one of the eight effects. The last slider determines the strength of the effect. The schematics also show a window of a participant of a slider. When participants move the slider (and thus the voice), the voice configuration updates one parameter in the voice configuration (in the image the ``speed'' dimension is changed).}
  \label{fig:architecture}
\end{figure}

To create a voice for a robot, we need an expressive voice creation tool that is fully parametrizable. Our solution is depicted in Figure \ref{fig:architecture}. 
Overall, the architecture changes the voice of a Text-To-Speech (TTS) model, changes the speaking speed, and passes the resulting audio to a rack of effects. Participants use sliders to adjust the model parameters, thus changing the voice.

The first five sliders modify the voice of the speaker of a TTS model. We modified the state-of-the-art TTS model \textit{VITS} \cite{vits} trained on the VCTK dataset \cite{yamagishi2019vctk} so that it can be used to directly modify the voice representation (speaker embedding). We performed a Principal Component Analysis (PCA) on all 110 speaker embeddings of the same dataset. We use the first five PCA components, which seem to capture sufficient variation in the human voice, as voice sliders (see Supplementary Materials \ref{sup_voice_dimension_reduction} for further details). These dimensions have no direct interpretation, but they correspond to intuitive vocal features such as gender, speaking speed, and voice timbre. We perform reverse PCA to obtain a speaker embedding based on the PCA dimensions. For maximum expressivity and minimum distortion, the range is constrained to approximately four standard deviations in all dimensions.
 
Since the variability in speaking speed in natural human speech is rather limited and the PCA dimensions by themselves did not provide enough variability in terms of duration, we added a sixth slider that can parametrically change the speaking speed ranging from 46\% to 153\% of the original speed using Parselmouth \cite{parselmouth}, a Python wrapper for Praat \cite{praat}.

Since we could not find a suitable dataset of robot voices, we trained our TTS system on natural speech (VCTK). This means that the TTS model mainly produces naturalistic human voices and does not create robotic-like sounds. Therefore, we added sliders to apply robotic audio effects. Here we combined modern TTS with traditional signal processing techniques. We implemented eight different effects commonly used to create robotic voices: changing the pitch, decreasing synthesis quality, applying a timeshift, using a vocoder, or applying one of four different flanger configurations to the audio. We implemented an effect rack using the Librosa Python library \cite{mcfee2015librosa}, which applies the effects in a sequential order. To avoid a strong mixture of voice effects, participants used a seventh slider to pick one of the eight effects and used an eighth slider to adjust the strength of the effect. The overall amplitude of the effects was manually normalized such that each effect would be approximately equally salient. The slider positions are linearly spaced (with a resolution of 16 positions) to make the synthesis computationally efficient.
\noindent We used the following types of effects. Note that the exact parameters for the effects are described in the Supplementary Materials \ref{sup_effects} and implementation is provided in the code repository (\url{https://robotvoice.s3.amazonaws.com/code.zip}):
\begin{itemize} 
\item \textbf{Pitch.} We enhanced the signal with two transposed audio tracks, where one was transposed five semitones up and the other transposed five semitones down. By doing so, the intonation pattern of the voice gets obscured, resulting in an unnatural voice. Further, both transposed signals are a minor seventh apart, which is generally considered a rather dissonant interval in Western music perception \cite{costa2000psychological}. As such, additional tension in the voice is induced. The corresponding slider in our experiments allows us to control the ratio between the non-transposed and transposed signals.

\item \textbf{Synthesis quality.} Older text-to-speech systems are poor at phase reconstruction, which results in audible artifacts that sound "robotic." To emulate this poor reconstruction, we transformed the signal to the frequency domain using a short-time Fourier Transform. We then reconstructed it using an inverse short-time Fourier Transform but with randomly initialized phase estimates. Our implementation utilized Librosa's Griffin-Lim algorithm \cite{mcfee2015librosa} without executing phase approximation.

\item \textbf{Timeshift.} To facilitate the creation of more "fuzzy" sounds, we also provided the option to blend the original voice with a slightly time-shifted version of the original signal. By doing so, the warmth and resonance of a natural voice gets veiled. To obtain this effect, the original signal was delayed for a few milliseconds, and the time-shifted signal was combined with the original signal.

\item \textbf{Vocoder.} Vocoder effects are frequently used to create robotic voices \cite{ramirez2017robotization}. We used the speech signal as a modulator for a carrier signal. By fixing that carrier signal to a certain frequency, the resulting voice sounds monotone and mechanical. Our pipeline makes use of TAL Vocoder\footnote{\url{https://tal-software.com/products/tal-vocoder}}, a publicly available VST implementation, which we included into our codebase using Pedalboard\footnote{\url{https://github.com/spotify/pedalboard}}.

\item \textbf{Flanger.} We incorporated a flanger, an audio effect that imparts a more synthetic quality to the sound. The flanger effect is achieved by combining a signal with a delayed version of itself where the delay time is modulated by a low-frequency oscillator. This addition offers an avenue to offset the voice's natural tone. We made four distinct flanger variants available, each producing a unique auditory experience.

\end{itemize}

\noindent For each robot we randomly selected a sentence from the 720 phonetically balanced and semantically neutral Harvard sentences \cite{harvardsentences}.

\subsection{Robot attributes proposed in the literature}
\label{attributes_literature}
To obtain a long list of attributes proposed for robots in literature we included labels from the ``Big Five'' personality model \cite{trapnell_extension_1990,mccrae_introduction_1992,gosling_TIPI_2003,rammstedt_bfi-10_2007}, from the Godspeed questionnaires that focuses on social robots \cite{bartneck_godspeed_2009}, from the relevant dimensions that V\"olkel et al. identified for voice assistants \cite{voelkel_developing_2020}, and from the AttrakDiff questionnaire that focuses on user experience in general \cite{Hassenzahl_attrakdiff:_2003}.
Furthermore, we added three adjectives signifying demographic features, namely ``young'', ``male'' and ``female'' to specify age and gender. We also added the word ``animallike'' because our collection of robots contains many artificial pets and animal-inspired robots.
This yielded 260 unique attributes (see Supplementary Tables \ref{stab:260_attributes_1}--\ref{stab:260_attributes_4} for the full list). While this list clearly does not capture all possible attributes ever mentioned for robots, it covers the most widely used attributes in the literature.

\subsection{Participants and experiments}
Overall, we recruited 2,505 participants. Participants were recruited from Prolific and provided informed consent under an approved protocol, and data was collected anonymously, with participants identified only by their prolific ID in order to enable compensation. Participants earned 9 pounds per hour, had a minimum age of 18 years, had to live and be born in the UK, had to speak English as their first language, and had to have been raised monolingually. See Supplementary Materials \ref{sup_participants} for additional demographic information about the participants and the number of participants in each experiment. All experiments were implemented with PsyNet \footnote{Psynet is available here: \url{https://www.psynet.dev/}. PsyNet \cite{harrison2020gsp} relies on the open-source platform Dallinger (\url{https://dallinger.readthedocs.io/})}, which is a framework for large-scale behavioral research. (see Supplementary Materials \ref{sup_implementation}) If audio was played in the experiment, we made sure participants were wearing headphones \cite{woods2017headphone}. 
If the experiment involved a lot of text (see Supplementary Table \ref{stab:table-experiments}), we tested English proficiency by an objective test that goes beyond their self-report (see Supplementary Materials \ref{sup_prescreen}).

\section{Results}
\subsection{Human-in-the-Loop Voice Creation}
\label{gsp_results}
\noindent 803 UK participants engaged in a GSP experiment (see Supplementary Materials \ref{sup_participants} for demographic information).
In the study, participants were tasked with tailoring voices to 175 robot images (each participant visits 20 different robots) by manipulating one slider at a time in order to tweak vocal parameters to best match the voice with the robot's appearance (as depicted in Figure  \ref{fig:hitl_paradigms}A, see Supplementary Materials \ref{sup_gsp_instructions} for instructions).  Initially, all vocal parameters were uniformly randomized with the possible range values (see Methods, Section \ref{methods_voice}). We then presented the slider to five participants, and the median of their responses was carried forward to the subsequent iteration (in the case of the effects slider, we picked the majority vote, see Supplementary Materials \ref{sup_gsp_aggregation} for further justification of the choice of median). In the next iteration, the aggregated parameters from the previous generation are propagated and a new group of five participants are recruited to control a different voice dimension. The idea is that the subjective match of voices to images gradually increases over iterations \cite{harrison2020gsp}. The sequence in which the dimensions were altered was shuffled for each robot. The experiment concluded after 48 hours, during which time 70 images underwent 15 iterations, and 105 images experienced 16 iterations. Consequently, each of the eight dimensions was visited approximately twice.

\begin{figure}[ht!]
  \includegraphics{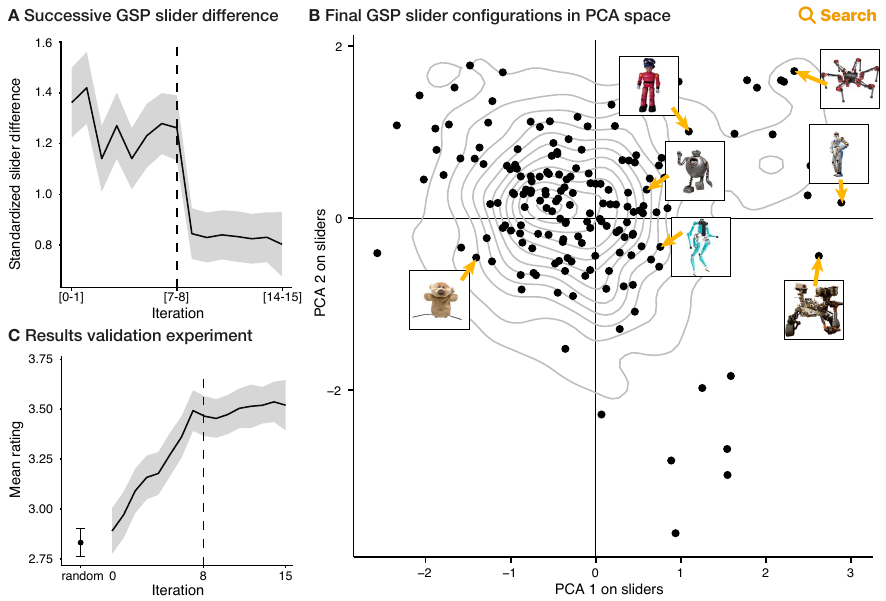}
  \caption{GSP results. \textbf{A} Standardized difference between successive slider configurations. \textbf{B} PCA on all slider configurations from all iterations. The gray kernel density estimate indicates the distribution of all slider configurations in PCA space. The black points are the final slider configurations. \textbf{C} Mean ratings as a function of the iterations and a random voice. Shaded areas are confidence intervals.}
  \Description{The figure is divided into three panels. A: The first panel shows the standardized difference between successive slider configurations. The subfigure shows that the slider difference goes down over the course of iterations and stabilizes after each dimension is visited once. B: Shows a 2D projection of the slider configurations using PCA. The figure shows that the final voices are quite different from each other, but that there are clusters of robots that look alike. C: Line plot of the mean ratings as a function of the iterations and a random voice as a point plot. The mean rating for random is the lowest. The rating for iteration 0 is similar. The average rating increases as a function of iterations, but the increase becomes less once each dimension has been visited once.}
  \label{fig:gsp_results}
\end{figure}

Figure \ref{fig:gsp_results}A shows that the standardized slider difference between consecutive iterations within a chain decreases over the course of iterations. This means that participants move the sliders to a lesser extent at later iterations, indicating convergence. In particular, there was a significant difference between the first and last iteration (Wilcoxon signed rank test:  \textit{V} = 11277.0, \textit{n} = 175,  \textit{p} < 0.001, \textit{r} = .43, this and all future tests are Bonferroni corrected for multiple comparisons) but we did not find a significant difference between the last iterations to the six iterations preceding it. The slider difference drops after all eight dimensions have been visited once, which is in line with previous studies \cite{vanrijn2022voiceme, vanrijn21_gsp_gst, harrison2020gsp}. The development over iterations can be listened to online: \url{https://robotvoice.s3.amazonaws.com/iterations.html}.

To visualize the proximity of the matched robot voices to each other, we performed a PCA on the standardized slider positions of all stimuli in the experiment. Figure \ref{fig:gsp_results}B depicts the first two principal components and shows the distribution of all slider configurations using a kernel density estimate (gray lines). The initial robot voice configurations are uniformly sampled from the sliders but occupy distinct slider positions at the end of the experiment. For example, the spider-like robots in the upper right corner or the toy-like robots in the top left corner of the plot group together in slider space (i.e., they received similar voices in the final iteration). The final voices can be explored interactively using the online visualization: \url{https://robotvoice.s3.amazonaws.com/explore.html}.

In order to validate whether the voice and robot match improves over time, we recruited a separate group of participants (\textit{N} = 142) that rated how well the voice matches the robot (see Supplementary Materials \ref{sup_validation_instructions}). 
This experiment comprised 2,730 stimuli. All stimuli were generated in the GSP process with three additional random voices per robot. There were about 4.9 average ratings per stimulus. Overall, we had 13,597 human judgments in this experiment.  
As depicted in Figure \ref{fig:gsp_results}C, the average match increases over the course of iterations. 
In particular, the average of the last three iterations was significantly larger than the first three iterations (Wilcoxon signed rank test:  \textit{V} = 1813.5, \textit{n} = 175,  \textit{p} < 0.001, \textit{r} = .64). In addition, the increase in rating over iterations reduces after each dimension is visited approximately once. For example, we did not find a significant difference between the average of iterations 8-10 and the average of iterations 13-15 (Wilcoxon signed rank test:  \textit{V} = 6321.5, \textit{n} = 175, \textit{p} = 0.018, \textit{r} = .10).

\subsection{Open-ended Labeling}
\label{step_results}
What are the semantic labels that determine robot appearance and voice characteristics? To answer this we used STEP-Tag \cite{marjieh_words_2022}, a recently developed elicitation method to elicit labels from stimuli.
We recruited two new groups of participants to annotate the obtained final robot voices and the original images (\textit{N} = 59 and \textit{N} = 73 respectively). Each robot is sequentially annotated by 10 participants (see Supplementary Materials \ref{sup_step_instructions} for the participants' instructions). The process is adaptive: one participant provides annotation and subsequent participants rate it, flag it (in case they think it is inappropriate), or suggest their own annotation (Figure \ref{fig:hitl_paradigms}B). 
To facilitate convergence and avoid spelling variants and duplicate tags, participants can see words that start with the same letters while typing and can select them if they find them appropriate. The proposed words are either tags provided by other participants or the 260 dimensions proposed in the aforementioned literature (see Methods, Section \ref{attributes_literature}). 

\begin{figure}[ht!]
  \includegraphics{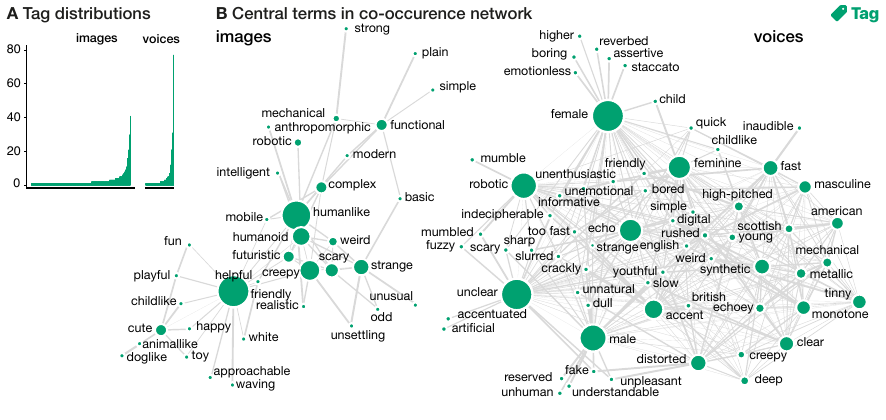}
  \caption{STEP-Tag results. \textbf{A} Raw occurrence of single labels for the 175 images and 175 voices. \textbf{B} Co-occurrence networks between provided tags per modality. Tags with a co-occurrence below 4 are pruned to remove words that are rarely used. The size of the nodes indicates the degree. Networks are created using Gephi \cite{gephi}.}
  \Description{A plot with two panels showing the STEP-Tag results. A: shows the distribution of tags in the image and voice modality. Participants a larger number of different words to describe their experience compared to the voice. In the voice dimension, the vocabulary is thus smaller, but the same words are used more frequently. B: Co-occurrence networks for the image and voice dimension. Overall the image network contains fewer items and is less densely connected than the network of the voice modality.}
  \label{fig:step_results}
\end{figure}

As depicted in Figure \ref{fig:step_results}A, the vocabulary used to describe the 175 images of robots is generally larger than those of 175 voices (765 and 217 unique tags for the image and voice modalities, respectively). Also, the same labels are used more frequently for the voice compared to the image modality (mean occurrence of 5.4 and 2.5 for the voice and image modalities, respectively).

To investigate which terms are particularly relevant, we visualize the co-occurrence network for each modality in Figure \ref{fig:step_results}B using a network analysis \cite{marjieh_words_2022}. The nodes are tags proposed in the STEP-Tag process and the edges indicate if they co-occur within the same robot with other tags. Those tags that have many connections to other tags -- indicated by the larger dots -- are likely to be relevant descriptors. In the co-occurrence network, terms that are semantically similar are often located near each other, such as 'animallike' and 'doglike' in the image modality. However, this isn't always the case, as terms that are semantically related do not necessarily appear together if they are inapplicable to a significant number of robots.

Overall, we observed that tags in the voice modality are more interconnected (average degree: image = 3.8, voice = 11.3), suggesting that a relatively small number of recurring labels frequently appear together. This observation aligns with what is shown in Figure \ref{fig:step_results}A. This pattern can be partially attributed to the challenge of identifying vocal properties compared to image attributes. Voice representations might be less easily described in words, or more ambiguous overall, leading to greater overlap in semantic labels.

Interestingly, while our approach is open-ended (e.g., we don't use post-processing and involve lay participants), many central terms overlap with those commonly mentioned in literature such as ``friendly'', ``humanlike'' or ``female'' (see, for example, \cite{fussell:et:al:2008} or \cite{powers2006advisor}). Figure \ref{fig:step_results}B furthermore reveals that while some impressions are modality-specific (e.g. ``high-pitched'', ``echo'', ``accent''), the majority of terms proposed by the participants reflect general impressions of the robot (e.g., ``weird'', ``cute'', ``robotic'' and ``friendly'') and are not modality-specific.  However, other features differ across the two modalities. For example, the biological sex or the age of the speaker is an important category in voices, whereas for the distinction between ``animallike'', ``humanlike'', and ``robotic''/``mechanical'' seems more important in the case of images.
Furthermore, the participants came up with terms for the voices that refer to communication qualities, such as ``inaudible'' or ``informative'', and the communication style, such as ``assertive'' and ``unenthusiastic''. Obviously, the participants were able to produce voices that complemented the visual impression of the robots by assigning additional attributes to them via the voice modality. The observation that participants used terms related to communication qualities and styles when judging voices, but not when viewing static images of robots, highlights the complementary of different sensory modalities, such as visual and auditory.

\subsection{Rate Robots along Perceptual Dimensions}
\label{dense_results}

To understand how the perceptual dimensions in the literature and the one from the STEP-Tag procedure describe each of the robots, we performed another experiment. Here a new set of 543 participants was recruited to rate all robots across a select set of dimensions. 
As a result, we chose familiar dimensions in order to capture labels that existed in the literature as well as labels that were perceptually salient to participants. We selected 40 attributes in order to have enough ratings per participant. 
Specifically, we selected the 26 dimensions that overlap between the list of 260 labels from previous literature (see Supplementary Table \ref{stab:260_attributes_1}--\ref{stab:260_attributes_4}) and our STEP-Tag results. The remaining 14 dimensions are the 7 perceptually most salient features (based on STEP-Tag) in each of the modalities. Supplementary Table \ref{stab:40_dimensions} specify the 40 dimensions and their sources.

We recruited separate groups of participants to rate these 40 perceptual labels in the image and voice modalities (\textit{N} = 298 and \textit{N} = 245 respectively, see Supplementary Materials \ref{sup_used_labels}). Each participant rated the robot image or robot voice on 5 randomly selected dimensions using sliders that snap to 5 positions. On average, each stimulus and dimension was rated 7.5 times for the images and 6.1 times for the voices (see Supplementary Materials \ref{sup_dense_instructions} for the experiments' instructions). Overall, the ratings were reliable for both experiments: the split-half reliability for images was \textit{r} = 0.65 and \textit{r} = 0.61 for the voices. To compare the consistency of the dense rating results with the STEP-Tag results in the previous experiments, we correlated STEP-Tag ratings with the dense ratings for the labels that occur in both datasets. As shown in Supplementary Materials \ref{sup_step_consistency}, there is a diagonal for most terms indicating that there is a strong correlation between the number of stars a label received in STEP-Tag and the average rating it received in the dense rating experiment (mean diagonal: \textit{r} = .31 and \textit{r} = .24, off-diagonal: \textit{r} = 0.11 and \textit{r} = 0.10 for images and voices respectively). 

\begin{figure}
  \includegraphics{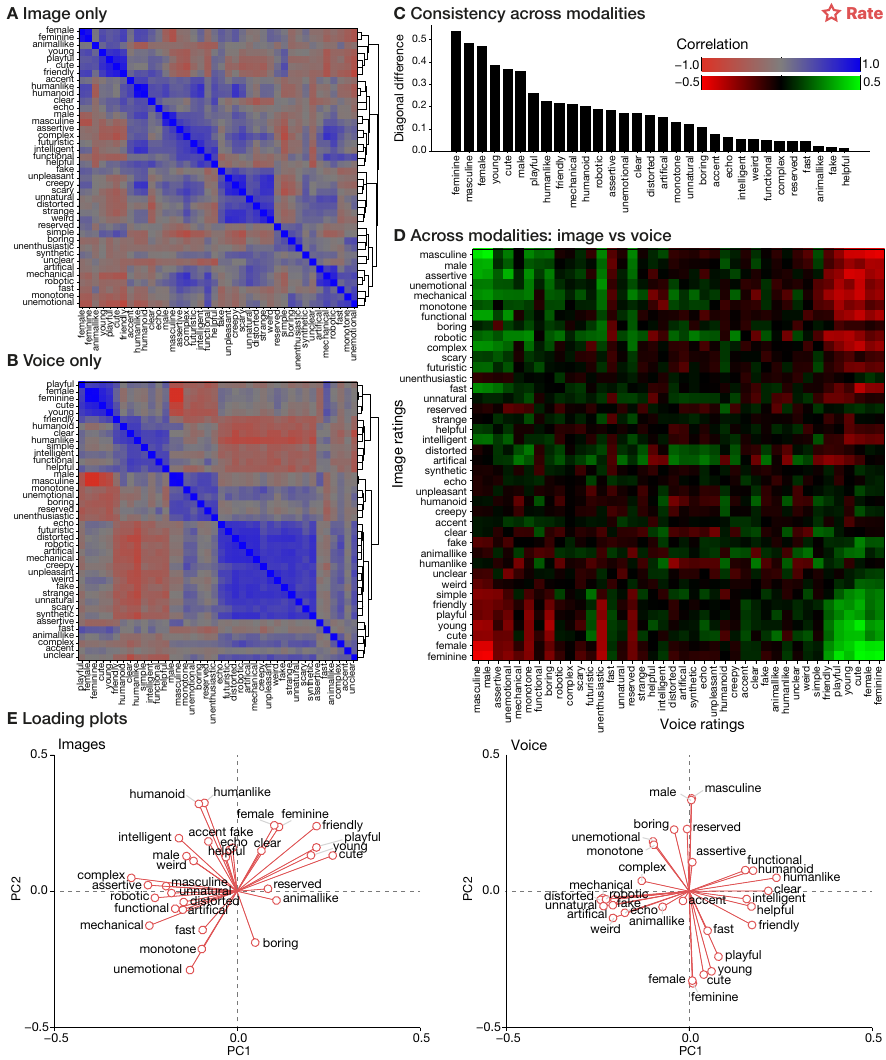}
  \caption{Correlations between ratings along dimensions. Correlation across dimensions for \textbf{A} images and \textbf{B} matched voices. Correlation matrices are sorted by the order in the dendrogram obtained via agglomerative clustering. \textbf{C} Most consistently rated dimensions across both modalities. The diagonal difference is the difference in correlation between the diagonal and the mean correlation of the rest of the row. \textbf{D} Correlation across both modalities. The correlation matrix is sorted by mean correlation for the most consistently rated dimension ``feminine''. \textbf{E} Loading plots for both modalities. PCA components were obtained separately for the data of the correlation matrices in panels A (left) and B (right).}
  \Description{Figure showing the dense rating results. A: Shows the correlation matrix in the image modality. B:  Correlation matrix in the voice modality. The correlation matrix in the voice modality has more of a block structure compared to the image modality. C: Comparing the correlation between the ratings across modalities. The following dimensions correlate strongest across modalities: ``feminine'', ``masculine'', ``female'', ``young'', ``cute'', ``male'', ``playful'', ``humanlike'', ``friendly''. D: Correlation matrix across modalities sorted by the strongest dimension across modalities (``feminine''). We find strong correlations between gender and other attributes, such as ``male'' and ``assertive'' or ``female'' and ``cute''. E: Loading plot for image and voice modality. In the voice modality, the first principal component primarily captures the contrast between “humanlike” and "robotic,” while the second dimension focuses on the male-female dichotomy. In the image modality, a similar contrasting pattern is observed between “humanlike” and “robotic” features, but here the emphasis is on terms related to automaticity, such as “fast,” “monotone,” and “unemotional,” as opposed to terms like “playful,” “friendly,” and “cute”.}
  \label{fig:dense_results}
\end{figure}

Figure \ref{fig:dense_results}A shows the correlations between the dimensions for the image modality (i.e., a correlation between average rating per stimulus between all dimensions). Generally, terms with similar meanings, such as "female" and "feminine," show strong positive correlations, while antonyms like "clear" and "unclear" display strong negative correlations. The matrix reveals an additional pattern: participants tend to associate female robots with labels like "young," "playful," "cute," and "friendly," while male robots are linked with traits such as "assertive," "functional," "complex," and "intelligent." These observations align with previous literature \cite{powers2006advisor}, which suggests that societal stereotypes influence how robots are perceived. 

For the voice, the correlation matrix shows a more consistent structure (Figure \ref{fig:dense_results}B): The largest cluster contains dimensions like ``creepy'', ``unpleasant'', ``mechanical'', and ``robotic''. Also ``female'' is associated with a ``young'' and ``cute'' voice (consistent with previous literature) \cite{Cambre:Kulkarni:2019}, but not with a ``friendly'' voice. Instead, a new cluster emerges for ``friendly'', ``helpful'', ``clear'', and ``intelligent'' voices. This suggests that voice modality presents a much harder challenge in terms of providing labels. Specifically, voices cluster to a smaller number of interconnected terms (consistent also with the usage of smaller vocabulary in Figure \ref{fig:step_results}A).

To investigate the robustness of our findings, we run the dense rating experiment on 175 new images from the ABOT dataset \cite{Phillips2018} (see Supplementary Materials \ref{sup_new_robot_dataset}) and on 175 randomly created voices using our voice tool. We found strong correlations between the two image (\textit{r} = .85) and two voice datasets (\textit{r} = .91, see Supplementary Materials \ref{sup_dense_generalizability}). These findings indicate that the obtained correlations across the terms are robust across datasets.

We also investigated the correlations of the dimensions across the modalities. Generally, the correlations were lower, indicating that the association between the dimensions across the modalities is weaker (e.g., a masculine robot does not necessarily become a male-sounding voice, see Figure \ref{fig:dense_results}C for the correlation between terms across modalities). Furthermore, as depicted in Figure \ref{fig:dense_results}D, the diagonals between the dimensions were much weaker or entirely vanished for certain dimensions for example for terms like ``humanoid'' or ``unpleasant''(see Supplementary Materials \ref{sup_consistency_across_modalities} for a correlation matrix sorted by the strength of the diagonal). This indicates that the same labels are not consistently used across modalities, e.g. a ``fast'' voice does not mean that the image of the robot looks ``fast'' too. The dimensions that are best preserved across modalities are dimensions like ``feminine'', ``young'', and ``cute'' (Figure \ref{fig:dense_results}C).

In Figure \ref{fig:dense_results}D, we can see that there is a large overlap between associations from images to voices as well as from voices to images (e.g., male voices are associated with mechanical robots, and vice versa). However, this relationship is not always bidirectional; for example, assertive robots are associated with male voices (\textit{r} = 0.32), but male robots are not really associated with assertive voices (\textit{r} = 0.09). Further comparisons between the modalities can be made via the interactive visualization: \url{https://robotvoice.s3.amazonaws.com/compare.html}.

Figure \ref{fig:dense_results}E displays the factor loadings for consistent dimensions across modalities as they relate to the first two principal components in the data from the correlation matrices shown in Figures \ref{fig:dense_results}A and \ref{fig:dense_results}B. A high loading indicates a strong alignment between a specific word and the PCA dimensions. In the voice modality, the first principal component primarily captures the contrast between ``humanlike'' and "robotic,'' while the second dimension focuses on the male-female dichotomy. In the image modality, a similar contrasting pattern is observed between ``humanlike'' and ``robotic'' features, but here the emphasis is on terms related to automaticity, such as ``fast,'' ``monotone,'' and ``unemotional,'' as opposed to terms like ``playful,'' ``friendly,'' and ``cute''. The gender dichotomy is somewhat less pronounced here. 

\begin{figure}
  \includegraphics{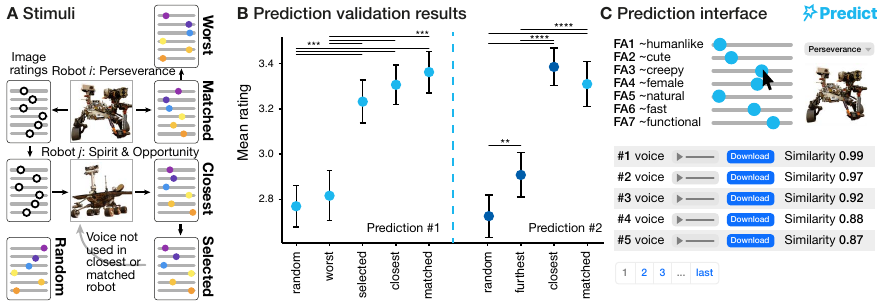}
  \caption{Prediction of a voice. \textbf{A} Schematics of the procedure to select stimuli for the prediction validation experiment. \textbf{B} Results of the prediction validation experiment. Matched receives the highest score followed by the closest and selected voice configuration. The worst and random voices receive the lowest scores and are significantly lower than the matched, closest, and selected voice configurations. *** indicates that the paired Wilcoxon signed rank test was significant ($p<0.001$). \textbf{C} Schematics of the prediction interface.}
  \Description{Figure showing the prediction results and interface. A: Schematics of the procedure to select stimuli for the prediction validation experiment. For a given robot, we look up the matched voice and the perceptual ratings in the image modality. We then look for the closest robot and look up its voice. We now look for the voice which is most near to this voice. In addition, we also add the worst voice (i.e., the voice with the largest cosine distance to the matched voice) and a random voice. B: Results of the validation experiment. We show that there is not a significant difference between the matched, closest, and selected voices, but all three have a significantly higher score than the worst or random voice. C: Schematics of the prediction interface. The subfigure shows seven sliders controlling latent dimensions in the image space. Changing the sliders will select the closest robot image. Alternatively, the user can select the closest robot from the dropdown list. Below the sliders and the image of the selected robot, the closest voices are listed. The voices can be downloaded and used in an application directly.}
  \label{fig:prediction_results}
\end{figure}
\subsection{Predict Voices based on Labels}
\label{prediction_results}

Finally, we wanted to learn if our results can be used in practice for engineers who want to fit a voice to a robot. To test whether we can predict the voice of a robot based on the image ratings in the dense rating experiment we recruited a new group of  (\textit{N} = 94) participants. We assessed whether the obtained perceptual dimensions can be used to propose a well-matched voice to an unseen robot. For each robot image $i$, we provide five different combinations of an image and a voice (Figure \ref{fig:prediction_results}A): As ground truth we included the original matched voice of robot $i$ (\textit{matched}).
To see how well we can use verbal descriptors to perform voice prediction, we searched for the robot $i$ with perceptual image rating across the 40 dimensions and found the closest robot $j$ (\textit{closest}, i.e., with the highest cosine similarity, e.g., the Perseverance robot is closest to the Spirit \& Opportunity robot). We then used the voice of the $j$ in the final iteration of the GSP experiment. To test robustness, we also searched for robot $j$ for the GSP slider configuration and selected the closest voice in slider space, which did not occur in any iteration for robot $i$ and $j$ (\textit{selected}). As a negative reference, based on the slider configuration of the matched robot $i$ we searched for the worst slider combination (\textit{worst}, i.e., which is maximally dissimilar in cosine similarity). Finally, we also included a random voice configuration (\textit{random}). The interface of the prediction experiment was identical to the GSP validation experiment (see Supplementary Materials \ref{sup_prediction_instructions}). We had 875 stimuli and 7,444 human judgments overall, and each stimulus received an average of 8.5 ratings.

Consistent with the validation of the GSP voices, the random voice received the lowest voice match score and the final voice the highest match score (Figure \ref{fig:prediction_results}B, left panel). While the closest and selected voices received a slightly lower match rating, we did not find a significant difference there (Wilcoxon signed rank test:  \textit{V} = 6879.0, \textit{n} = 175, \textit{p} = 0.47, \textit{r} = .07). However, the matched, closest, and selected voices were all significantly better than the worst or random voices (\textit{p} < 0.001 in all cases), which both have much lower ratings. Thus, this shows that while the predicted voices (closest and selected) were all better matches than a random voice, they were not significantly worse than the matched voice. This trend is not only visible when averaging over all participants, but also on a single-participant basis (see Supplementary Materials \ref{sup_prediction_by_participant}).

To assess if our findings also extrapolate to other datasets of robots, we run another prediction experiment (\textit{N} = 73). We wonder if the annotated features of the new robot can be used to match the voice based on the old data set's annotated features. In a real-world scenario where an engineer might have a new, unseen robot image and want to use our results for voice matching, this validation is crucial as it should show that even when using voices tailored to the old dataset and a matching model trained solely on the old dataset, we can still achieve accurate predictions with a new set of independently annotated images.
Thus, we looked up the closest robot in terms of its annotated features for each of the new 175 robots in the old set of matched robots (\textit{closest}).  As a reference, we also included the same matched voice and paired it with the directly matched old robot image (\textit{matched}). As a negative reference, we looked up the perceptually furthest robot in the old dataset and selected its voice (\textit{furthest}). We also add a random voice (\textit{random}). As shown in Figure \ref{fig:prediction_results}B (right panel), the closest and matched voices are all significantly better than the furthest and random voice (\textit{p} < 0.001 in all cases). While the closest voice received a slightly higher rating than the matched voice, this difference was not significant (Wilcoxon signed rank test:  \textit{V} = 8215.5, \textit{n} = 175, \textit{p} = 0.14, \textit{r} = .11). As in the previous prediction experiment, the furthest matched voice was slightly higher than random though both of them had low ratings overall. This is probably because random voices are uniformly sampled along the dimensions, leading in some cases to sample extreme values, which is not the case for the furthest or worst voices that were matched to a robot. This additional prediction experiment shows that our prediction also works for newly annotated robots from different datasets.

To facilitate a wide adaptation of the tool, we provide an interactive voice prediction tool online:  \url{https://robotvoice.s3.amazonaws.com/predict.html} (Figure \ref{fig:prediction_results}C). The tool allows one to select a robot from the 175 robots which is most similar to the robot that requires a new voice. The user can either search for the visually closest robot from the dropdown list or modify latent dimensions representing the 40 dimensions in vision (see Supplementary Materials \ref{sup_prediction_factor_analysis}). For example, slightly modifying the latent dimensions of the Zeno robot likely returns the voice created for the Milo robot because they look much alike. For the visual match, it will show the closest voices in the slider space. The voice configurations can be downloaded and can directly be integrated into applications.

\subsection{Control analysis}
In this paper, we have shown that by using human-in-the-loop approaches, participants develop voices matched to robots, identify attributes relevant to the perception of robots, and provide ratings along those dimensions that can be used to predict well-matched voices to entirely novel robots. A natural question that arises is whether neural networks can replace parts of the human pipeline. Here, we conduct two experiments on CLIP \cite{Radford2021CLIP} (see Supplementary Materials \ref{sup_dense_clip}) to avoid the human dense rating experiments since they involve many participants and thus are costly. We show that there is a moderate correlation between the cosine similarity across robot images computed on the human dense rating and the image embeddings (\textit{r} = .58 for the old 175 robots and \textit{r} = .51 for the new robots). This indicates that CLIP embeddings provide a fair proxy for the perceived similarity of robots. In a second analysis, we use CLIP to do the dense rating experiment (so each image receives probabilities of all 40 labels). We now compute the correlation across terms for the CLIP and human data. We find a similarly strong correlation across the CLIP results across datasets (\textit{r} =
.87) compared to the human results (\textit{r} = .85, see Supplementary Materials \ref{sup_dense_generalizability}), but the CLIP and human data are uncorrelated in both the old (\textit{r} = .04) and new image dataset (\textit{r} = .01). The results show that the correlational structure across the terms is consistent across the datasets, but varies greatly between CLIP and the human dense rating. While CLIP provides a proxy for the perceived similarity of robots, researchers and engineers should be cautious about blindly replacing parts of the pipeline by neural networks, as the models will introduce new biases in the annotation process.

To explore if we can actively reduce biases in our data, we rerun the STEP and dense rating experiment on the images but participants first take an implicit bias training (see Supplementary Materials \ref{sup_implicit_bias}). To measure implicit biases before and after the training, we use the widely-used implicit association test (IAT) \cite{Greenwald1998, Schnabel2008}. Based on previous literature, we expected that the training would have a short-term effect on participants' responses and reduce adverse stereotypes \cite{Lai2014}. We found that participants carefully read the implicit bias fact sheet (6/8 text-comprehension questions were answered correctly), but we did not measure a significant difference in bias before and after training.

Running the STEP experiment (N = 78), we found a large overlap in the used tags across STEP experiments with and without the training (see Supplementary Materials \ref{sup_step_bias}), and the frequency of the shared tags is strongly correlated (\textit{r} = .78). Moreover we found that if the data from the STEP experiment after the awareness training had been used to compile the list of 40 terms, only one term would have been replaced. These findings suggest that the STEP tag results remained largely unaffected by the training. When running the dense rating experiment (N = 202) (see Supplementary Materials \ref{sup_dense_bias}), we found that the correlation matrices in the dense rating experiment with and without training strongly correlate with each other (\textit{r} = .91). This indicates that while participants were aware of their implicit biases (see comprehension questions), they did not substantially change their responses.

\section{Discussion}
The present work provides a voice creation tool that can cover a wide range of robotic voices (Figure \ref{fig:intro}A). We used this tool in a human-in-the-loop approach (GSP) to create matched voices for 175 robots (Figure \ref{fig:intro}B), obtained a taxonomy using a human-in-the-loop open-ended labeling approach (STEP-Tag, Figure \ref{fig:intro}C), densely rated the attributes from the taxonomy for 175 robots (Figure \ref{fig:intro}D), and predicted suitable voices for new robots based on those perceptual dimensions (Figure \ref{fig:intro}E).

\subsection{Limitation and Future work}

Our paper primarily focused on conveying robot characteristics through manipulating the audio channel. To control for voice manipulation, participants were presented with static images. The way robots move can significantly affect human perception, and a wide range of literature illustrates how robots convey personality through body language, gestures, and facial expressions, as summarized in \cite{Janowski2022}. Future research can investigate how different use cases and scenarios of the same robot can affect the perceived appropriate voices. Another limitation is that the voices we used were matched with short, semantically neutral sentences, which might not generalize to longer textual content. Consequently, participants formed opinions based on limited information about the robot and its voice. An intriguing future direction for this research could include the employment of dynamic materials such as videos instead of static images, as well as the use of longer, semantically relevant spoken content. While such complexities are beyond the purview of our current study, which is focused on the vocal channel, our methodologies could serve as a foundation for more comprehensive studies into voice interactions in dynamic robot settings.

While we purposefully selected a neutral background for the robot to minimize contextual biases, it is essential to recognize that participants may have held varying perceptions of the robot's role, task, and target audience while adjusting voice dimensions. Empirical evidence indicates that factors beyond the robot's attributes, such as the task and user characteristics, significantly influence how it is perceived and how humans interact with it \cite{kuchenbrandt:et:al:2014}. The transition from a toy-like robot to a robot serving as a speech assistant, as highlighted by Aylett \cite{aylett2019right} using the example of the Cosmo robot, can result in a mismatch between the robot's function and its voice. Tags used by our participants to describe the robot, such as ``functional'' and ``helpful'', highlight the importance of its intended purposes and audience in addition to its audio-visual characteristics. Both the robot's visual appearance and the mental models it triggered in participants may have influenced voice modifications. To gain further insights into these factors, conducting additional experiments with systematic changes to the robot's visual context, aligned with its intended functions, could be valuable.

At the end of the Results section, we have alluded to the possibility of replacing a part of the human pipeline with neural networks. Here, we used CLIP \cite{Radford2021CLIP}, but in future research, it would be particularly interesting to do similar experiments on Large Language Models with vision grounding, such as Gemini, GPT-4V, or Bard, as they both have access to image and language data.

We based our voice creation tools on an English dataset, which, while diverse in including multiple dialects, did not allow us to explore the intricate relationship between culture and robot perception. This limitation applies to the user's cultural background and the culture the robot is intended to portray. Prior research \cite{haering:et:al:2014} demonstrated that a robot's social category membership, including culture, significantly influences how people perceive and interact with it. During the annotation process, our participants included tags related to English dialects like ``Scottish'' and ``American'', highlighting the relevance of group membership as a distinguishing characteristic. McGinn and Torre \cite{mcginn2019can} manipulated the accent of a robot's voice to investigate its impact on the formation of stereotypes. However, due to the heterogeneous background of the participants, their findings on the effect of accent manipulation were not consistent. Further research should, therefore, focus on the alleged cultural background of robots portrayed by their accent.

In a broader perspective, our study only involved monolingual English UK participants and future research should incorporate less ``WEIRD" participants (Western, Educated, Industrial, Rich, Democratic) \cite{barrett2020crosscultural, henrich2010weird} to uncover associations across perceptual dimensions in different cultures. Our approach is largely language-agnostic (it would solely require a TTS model trained on a different language) and thus can be applied to a variety of languages and cultures.

Finally, while the matched voices are significantly better than a random voice (Figure \ref{fig:prediction_results}B), the mean ratings for the matched voices (3.4) are not quite at ceiling performance (5.0). This can have multiple causes. One explanation is that the voice model is not expressive enough yet. The voice dimensions mainly capture aspects of the voice such as gender or sex (see Supplementary Materials \ref{sup_acoustic_correlates}). Future research can improve the parametrization of the latent voice dimensions to capture more expressive features of the voice. Another possible cause is that participants disagree about the voice properties associated with a robot. The split-half reliabilities in the rating experiments are high but there is still some disagreement across participants (dense: image \textit{r} = .68 and .53, voice \textit{r} = .65 and .48; prediction: \textit{r} = .56 and .53). This indicates that future research should investigate individual differences in the perception of robots.

\subsection{Ethical considerations}
Our tools have the potential to uncover implicit biases that carry over from human-human interactions to human-robot interactions. For instance, our study revealed that participants tended to use tags like ``playful'' and ``friendly'' when describing a female robot and voice, whereas ``assertive'' and ``functional'' were more commonly associated with a male robot's image. Furthermore, when describing a male voice, participants frequently employed tags like ``unemotional'' and ``reserved''.

To assess if the correlational structure obtained here generalizes to other datasets, we repeated the dense rating experiment on new image and voice data and found strong correlations across the new and old datasets (\textit{r} = .85 and \textit{r} = .91 respectively). While this shows that our findings are robust across datasets, it does not rule out the possibility that both datasets are biased in the same way. To quantify the effect of the human-in-the-loop approach on perceived biases, we took, as an example, perceived gender. Consistent with the observation by Perugia and colleagues, we observe an underrepresentation of perceived female robots in both image datasets \cite{Perugia2022} (13 \% for IEEE robots and 19 \% for ABOT, percent below the midpoint for scale; see Supplementary Materials \ref{sup_selection_bias}).

However, when taking random voice samples from the text-to-speech model, participants evaluate the perceived gender of samples as nearly balanced (50 \%). This is consistent with the dataset the model was trained on \cite{yamagishi2019vctk}, which was intended to contain a diverse set of voices. Importantly, the percentage of perceived females in the GSP-matched voices was similarly balanced (49 \%). This means that despite viewing an unbalanced dataset of images, the human-the-loop approach provided a much more balanced voice distribution.  
Future research can use more balanced sets of robot images or use GSP with a generative model to create images of robots, which would lead to the development of ``customizable robots'' as proposed by Schiebinger \cite{schiebinger:2019}.

Next to potential dataset bias, we explored how implicit biases of participants actively can be reduced. Before rerunning the STEP and dense rating experiments, participants undergo implicit bias training and take the implicit association test (IAT) \cite{Greenwald1998, Schnabel2008}. While participants carefully read the implicit bias fact sheet (6/8 text-comprehension questions were answered correctly), we did not measure a significant decrease in bias. As a consequence, the intervention did not substantially alter the responses. This may be explained by the fact that the effects of the training are short-lived and hence barely change the implicit biases \cite{Lai2014}. Future research can consider other interventions to reduce biases in the data. For example, a common theme in the development of recent Large Language Models is to perform an additional refinement on the model to suppress averse responses using human supervision \cite{openai2023gpt4}. We can perform an analog step for our approach, we can add a final post-processing step in which humans are asked to flag implicit biases.


Finally, we want to emphasize that the relationships across the terms we uncovered within and across two modalities are not causal, but are merely correlations. So, the fact that images of female-looking robots tend to be perceived as ``cute'' does not mean that they are cute because they are female (e.g., female-looking robots might have a more fluffy appearance on average, which makes them look cute). 

\subsection{General conclusion}
The relationship between a robot's voice and its impact on user perception is complex. The primary aim of this paper was to explore the impact of nuanced voice features on individual perceptions of a diverse array of robots. In contrast to prior research -- that used a small number of robots, existing audio samples, and limited behavioral testing -- we adopted a multi-method approach that included generative AI, human-in-the-loop computations, and voice prediction. Our voice generation tool combines state-of-the-art TTS with traditional signal processing. We used human-in-the-loop computations in two key points in the research program: to collectively navigate a space of voice dimensions and to provide open-ended labeling of images and voices. We complement human-in-the-loop experiments with extensive behavioral validation experiments (\textit{N} = 2,505). Our results demonstrated that participants consistently converged towards specific voice prototypes that either enhanced or aligned with the attributes associated with the static images of the robots. Our findings highlight the significant interplay between visual and auditory perceptions in shaping how humans perceive and attribute qualities to robotic entities. Furthermore, our study revealed that predicting a suitable voice for images of previously unseen robots is possible. This discovery can be interpreted as evidence that static visual cues alone may suffice to empower individuals to create voices that consistently convey the collective mental model of the respective robot. Using the perceptual dimensions we obtained for the set of robots, we could propose suitable voices to designers for new robots, as well as reveal and possibly suppress societal stereotypes underlying participants' choices.  They are of practical relevance for engineers who want to fit a voice that matches a robot. More broadly, our research demonstrates the synergy between cognitive science and machine learning in tackling engineering challenges, such as human-robot interaction.

\begin{acks}

\end{acks}

\bibliographystyle{ACM-Reference-Format}
\bibliography{RobotVoice}

\clearpage
\newpage
\appendix

\setcounter{figure}{0}
\makeatletter 
\renewcommand{\thefigure}{S\@arabic\c@figure}
\makeatother
\setcounter{table}{0}
\makeatletter 
\renewcommand{\thetable}{S\@arabic\c@table}
\renewcommand{\bibnumfmt}[1]{[S#1]}
\renewcommand{\citenumfont}[1]{S#1}
\makeatother

\section{Code and data availability}
\label{sup_code}
A view-only anonymous link is provided to the public, containing all the data collected for this project during the review stage \footnote{\textbf{Code and data:} \url{https://robotvoice.s3.amazonaws.com/supplementary_materials.zip}}. It includes the new human behavioral data, the computational experiments with machine learning models, and all the necessary analysis scripts for producing the results. Additionally, the repository includes the PsyNet source codes for reproducing the behavioral experiments. Finally, we present an interactive visualization \footnote{\textbf{Interactive plots:} \url{https://robotvoice.s3.amazonaws.com/index.html}} for exploring the created voices, the perceptual space of robots and for predicting new voices based on the perceptual dimensions. 

\section{Behavioral Studies}
\subsection{Participants}
\label{sup_participants}
Participants were recruited from Prolific\footnote{\url{https://www.prolific.co/}} and provided informed consent under an approved protocol. The median age was 38 (SD: 12.6, min: 18, max: 88). 61.4 \% of the participants identified themselves as male, 36.6 as female, 0.1 \% as non-binary and 0.1 \% preferred not to say. The highest level of formal education is high school for 23.3 \%, college for 34.8 \%, graduate school for 25.1 \%, and postgraduate school or higher for 16.7 \% of the participants (0.1 \% of the participants had no formal education). The exact number of participants for each of the 7 behavioral experiments is reported in Table \ref{stab:table-experiments}.

\label{behavioural-studies}
\begin{table}[h!]
  \caption{Behavioral experiment summary table.}
  \label{stab:table-experiments}
  \centering
  \begin{tabular}{ccccccccc}
    \toprule
    Modality & Paradigm & Total stimuli & \vtop{\hbox{\strut Trials per}\hbox{\strut participant}} 
    & Section & $N$ & Pre-screening & Type & Dur.\\
    \cmidrule(r){1-9}
    Voices + Images & GSP & 175 & 20 & \ref{gsp_results} & 803 & HT & M & 11.8 \\
    Voices + Images & Validation & 3,255 & 80 & \ref{gsp_results} & 142 & HT & M & 15.3 \\
    Images & STEP-Tag & 175 & 30 & \ref{step_results} & 73 & WV & M & 12.7 \\
    Images & STEP-Tag* & 175 & 30 & \ref{step_results} & 78 & WV & C & 19.9 \\
    Voices & STEP-Tag & 175 & 30 & \ref{step_results} & 59 & HT, WV & M & 12.0 \\
    Images & Rating & 175 & 60 & \ref{dense_results} & 298 & WV & M & 13.5 \\
    Images* & Rating & 175 & 60 & \ref{dense_results} & 202 & WV & C & 20.8 \\
    Voices & Rating & 175 & 60 & \ref{dense_results} & 245 & HT, WV & M & 14.1 \\
    New Images & Rating & 175 & 60 & \ref{dense_results} & 249 & WV & C & 14.5 \\
    Random Voices & Rating & 175 & 60 & \ref{dense_results} & 189 & HT, WV & C & 15.0\\
    Voices + Images & Prediction & 875 & 80 & \ref{prediction_results} & 94 & HT & M & 10.1 \\
    Voices + New images & Prediction new & 700 & 80 & \ref{prediction_results} & 73 & HT & C & 11.2 \\
    \bottomrule
  \end{tabular}
  \begin{tablenotes}
    \item \textit{Note.} $N$ denotes the number of participants included in the analysis; WV denotes the WikiVocab English proficiency pre-screening task \cite{vanrijn2023wikivocab}; HT denotes the headphone test \cite{woods2017headphone}. * means that before the main experiment, participants did implicit bias awareness training. Type: M denotes an experiment for the main results, C denotes a control experiment. Dur. denotes the median duration in minutes.
\end{tablenotes}
\end{table}

The median total durations of the experiments are typical for online experiments. The duration of the GSP experiments are similar to other experiments using GSP \cite{harrison2020gsp, vanrijn21_gsp_gst, vanrijn2022voiceme}.

\subsection{Implementation}
\label{sup_implementation}
All behavioral experiments were implemented using PsyNet framework \cite{harrison2020gsp}. PsyNet is a novel experiment design framework that builds on Dallinger (\url{https://dallinger.readthedocs.io/}) and allows for flexible specification of experiment timelines as well as providing support for a wide array of tasks across different modalities (visual, auditory, and audio-visual). Dallinger is a modern tool for experiment hosting and deployment that automates the process of participant recruitment and compensation by integrating cloud-based services such as Heroku\footnote{\url{https://www.heroku.com/}} with online crowd-sourcing platforms such as Prolific.  Participants interact with the experiment through their web browser, which in turn communicates with a backend Python server responsible for the experiment logic. As an advantage of using PsyNet, it offers native support for adaptive human-in-the-loop experiments.

\subsection{Pre-screening}
\label{sup_prescreen}
In order to collect high-quality data, pre-screening tasks were used to avoid low-quality participants and users who used bots to respond. We conduct the pre-screeners right before the main experiment.
If the pre-screening tasks are not completed, the experiment will be terminated early, but the participants will still be paid for their time (regardless of the outcome). Pre-screeners are additionally ensure  two main criteria for data quality, namely, a) to ensure that participants are wearing headphones and can hear audio b) that they are native speakers of the language. To do this, we implemented two tasks from previous literature. Namely, an English proficiency test (\cite{vanrijn2023wikivocab}) for experiments that relied on text; and a standardized headphone test (\cite{woods2017headphone} used for experiments involving audio. Table \ref{stab:table-experiments} provides details on which pre-screeners were used in each of the behavioral experiments.


\begin{figure}[ht]
  \centering
  \includegraphics[width=0.5\linewidth]{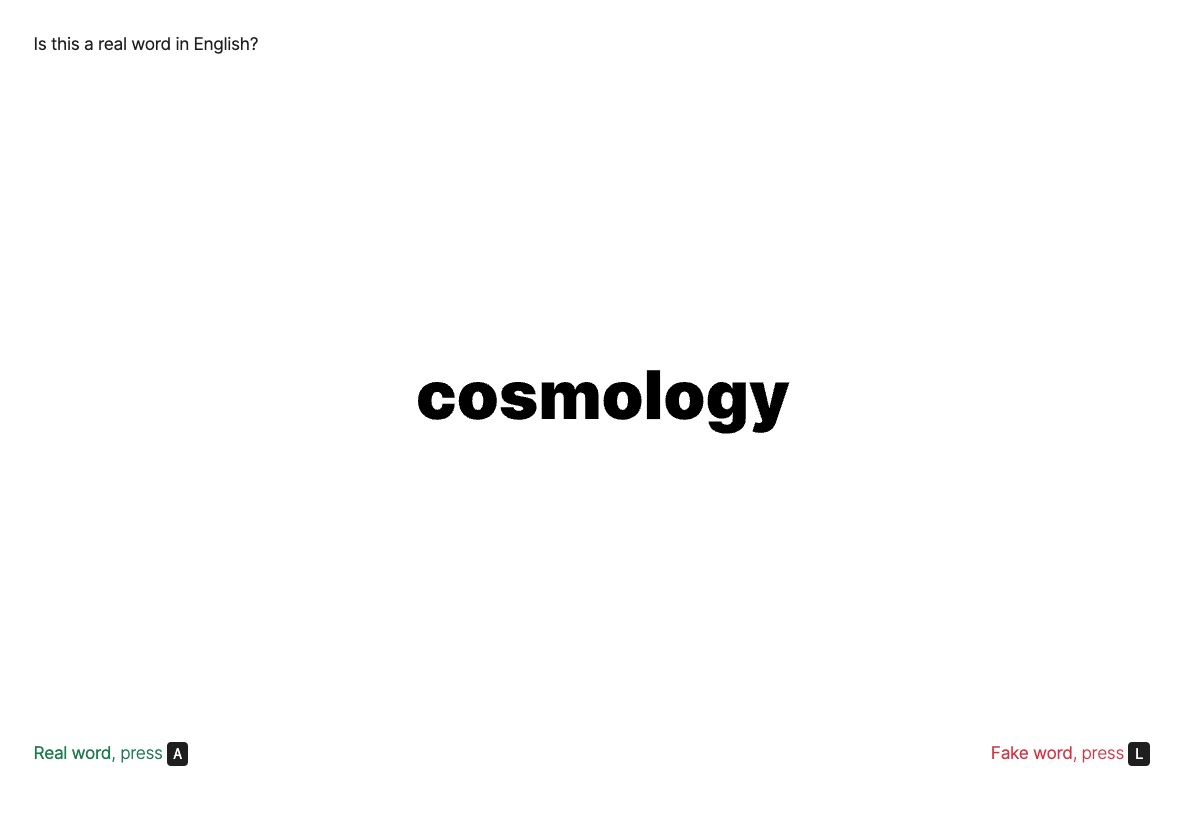}
  \caption{Example trial from the WikiVocab pre-screening task \cite{vanrijn2023wikivocab}.}
  \label{sfig:wikivocab}
\end{figure}

\noindent\textbf{English proficiency test}. To test participants' English proficiency we used the lexical decision task WikiVocab \cite{vanrijn2023wikivocab}. In each trial, we briefly present the participant (1 second) with either a real English word or a pseudo-word that does not exist. Participants were instructed to guess whether the word was real or not. They used dedicated keys on their keyboard to respond. A total of 30 trials (half of them being real words) were presented, and 25 of them needed to be correct for the participant to pass. For each batch of 30 trials, we randomly selected 15 real and 15 fake words. See \url{https://vocabtest.org/} for an implementation of the test. An example trial is shown in Figure \ref{sfig:wikivocab}.

\begin{figure}[ht]
  \centering
  \includegraphics[width=0.5\linewidth]{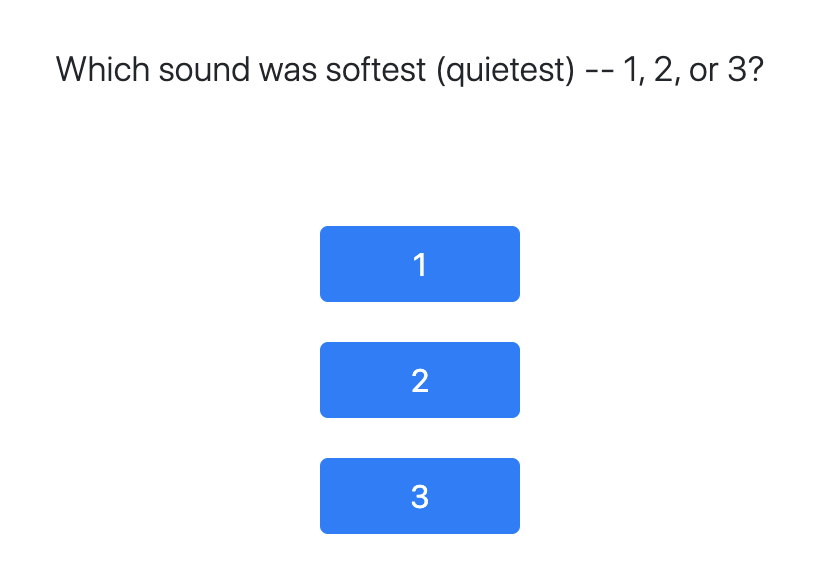}
  \caption{Example trial from the headphone pre-screening test \cite{woods2017headphone}.}
  \label{sfig:headphones}
\end{figure}

\noindent\textbf{Headphone test}. We used the headphone test developed by Wood et al. \cite{woods2017headphone}, which is used as a standard pre-screener for high-quality auditory psychophysics data-collection procedures \cite{milne2021online}. The test is designed to ensure that the participants are wearing headphones and are able to perceive subtle differences in volume. The task consists of a forced choice task, in which three consecutive tones are played, and the participant has to identify which of them is the quietest. Importantly, these tones exhibit a phase cancellation effect without headphones, making it difficult for non-headphone users to identify the quietest tone. To pass, participants had to answer 4 out of 6 trials correctly. An example trial is shown in Figure \ref{sfig:headphones}. 

\section{Methods}
\subsection{Selection of images}
\label{sup_image_selection}
\begin{figure}[ht]
  \centering
  \includegraphics{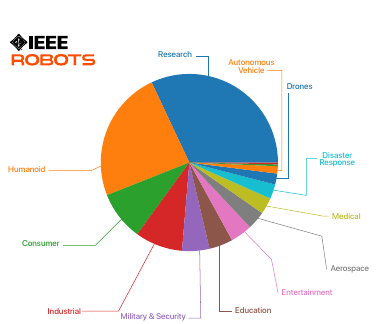}
  \caption{Distribution of different categories of robots used from IEEE Robots.}
  \label{sfig:image_selection}
\end{figure}

We selected a wide variety of different robots from the robot database IEEE Robots. The images span 14 categories as marked by the database (see Figure \ref{sfig:image_selection}). The majority of the images fall into the categories ``humanoid'' and ``research''.

Tables \ref{stab:robot_selection1}--\ref{stab:robot_selection2} shows examples of the images selected and edited for the experiment. The full list of stimuli is available at \url{https://s3.amazonaws.com/robotvoice/explore.html}.

\subsection{Selection of new set of images}
\label{sup_new_robot_dataset}
To assess the generalizability of our findings, we ran the dense rating on a new set of images and voices (see section \ref{sup_dense_generalizability}). We used the ABOT robot database\footnote{\url{https://www.abotdatabase.info/}} \cite{Phillips2018, Perugia2022} and obtained 167 new robot images from it after removing those that were already in the IEEE Robots selection.

We then added the following 7 robots that were used in a study by Mathur and Reichling \cite{Mathur2016} but not already included in the other two lists.
\begin{itemize}
    \item 3e A18 (Honda)\footnote{\url{https://global.honda/en/innovation/CES/2018/001.html}}
    \item 3e C18 (Honda)\footnote{\url{https://www.honda.com.au/news/2018/honda-3e-robot-concept}}
    \item aeo (Aeolus Robotics)\footnote{\url{https://aeolusbot.com/}}
    \item cruzr (Ubtech Robotics)\footnote{\url{https://ubtrobot.com/}}
    \item Jules (Hanson Robotics)\footnote{\url{https://www.hansonrobotics.com/jules/}}
    \item Actroid Repliee Q2 (Osaka University, Kokoro Co. Ltd)\footnote{\url{https://en.wikipedia.org/wiki/Actroid}}
    \item Tapia (MJI Robotics)\footnote{\url{http://mjirobotics.co.jp/en}}
\end{itemize}

Another robot that we added was Emotech's Olly \footnote{\url{https://www.indiegogo.com/projects/olly-the-first-home-robot-with-personality/}} because its abstract design is very different from that of the other robots and seemingly contrasts with the personality that its creators emphasized in advertising.

\subsection{Selection bias in images}
\label{sup_selection_bias}

\begin{figure}
    \centering
    \includegraphics{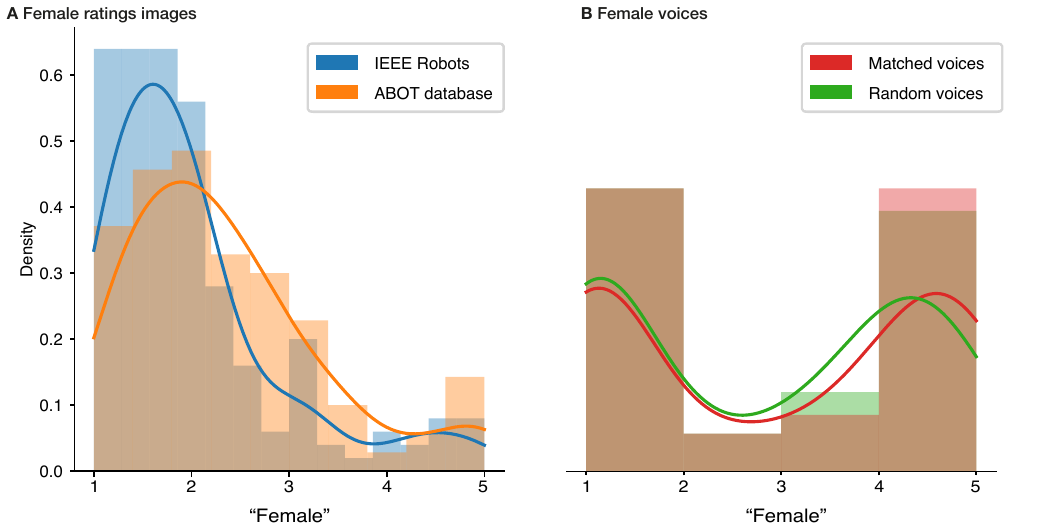}
    \caption{Distribution of ratings for attribute ``female'' in image (\textbf{A}) and voice datasets (\textbf{B}).}
    \label{sfig:dense_female}
\end{figure}

Consistent with previous literature, we find an underrepresentation of female robots in the ABOT database \cite{Perugia2022} and in the IEEE Robot database (see Figure \ref{sfig:dense_female}A). The initial random robot voices are gender-balanced (Figure \ref{sfig:dense_female}B). Interestingly, the matched voices to the biased sample of robot images are equally gender-balanced as the initial random voices.

\subsection{Latent voice dimensions}
\label{sup_voice_dimension_reduction}
\begin{figure}
    \centering
    \includegraphics[width=0.7\linewidth]{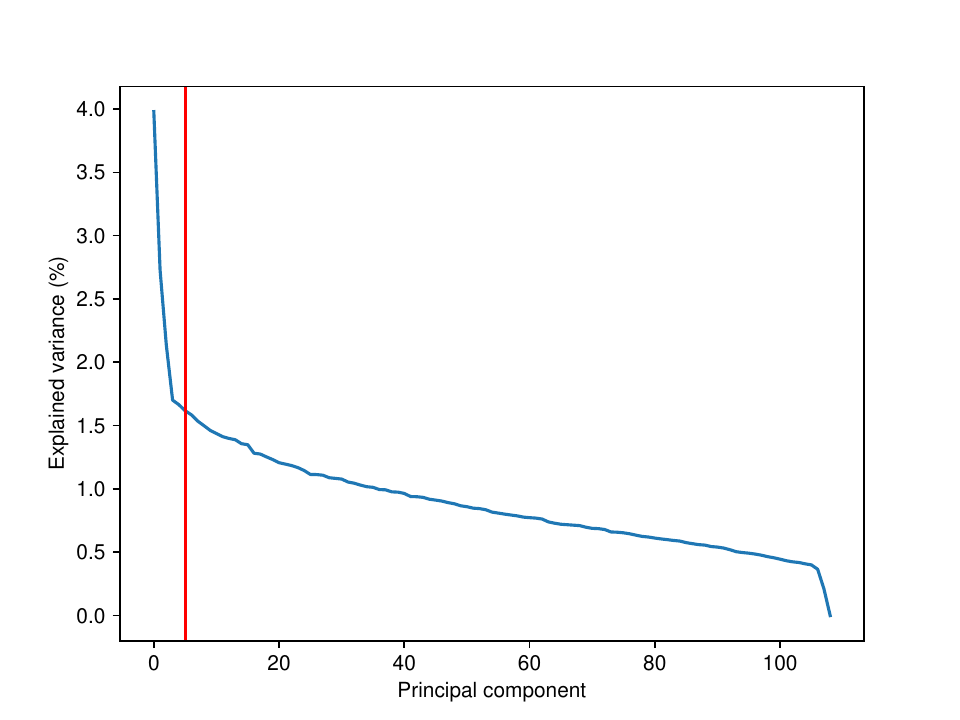}
    \caption{Explained variance by Principal Component Analysis.}
    \label{sfig:pca_explained_variance}
\end{figure}

In an initial set of pilots, we noted that using only the TTS system to create voices provides diverse human-like voices but does not allow us a way to create mechanical voices. Thus, adding effects was essential. Once we add effects, these effects can also account for some of the voice characteristics (e.g., pitch height, speech duration, voice roughness, etc.). This decreases the importance of having very high dimensional TTS representations.
We experimented with various dimension reduction algorithms. In particular, we experimented with supervised (PLS and CCA) and unsupervised dimension techniques (PCA). For PLS and CCA, we used the eGeMAPS feature set \cite{Eyben2016} as a supervision signal. Based on initial piloting, we found that the PCA voice dimensions had the best tradeoff between expressivity and distortion (high expressivity, low distortion). Across the dimension reduction methods, we found that the total amount of explained variance was comparable from method to method and was generally relatively small. This is compatible with previous studies in which the explained variance of the PCA on latent voice dimensions was low \cite{vanrijn2022voiceme} (e.g., 25.4 \% variance explained for ten dimensions). We reduced the total number of dimensions to five dimensions, capturing 12.2 \% of the variance (see Figure \ref{sfig:pca_explained_variance}). Note that the explained variance per dimension is high for the first few dimensions and decays slowly afterward, which makes the choice of five dimensions reasonable. Moreover, our pilot suggested that adding further dimensions did not improve voice expressivity qualitatively. Since most of the sliders in the experiment are voice dimensions (5/8), using a lower number of voice dimensions allows us to revisit dimensions more often in the GSP process and accelerate convergence. While we only use five dimensions, the dimensions clearly capture various aspects of the voice, such as sex and age (see Supplementary Materials \ref{sup_acoustic_correlates}).

\subsection{Voice Effects}
\label{sup_effects}
We implemented a set of audio effects to allow the creation of voices that sound more synthetic. For each effect, the slider in our GSP experiment steered the amount of effect in the resulting signal. The only exception here is the \emph{Timeshift} effect, where the slider did not control the amount of the effect but the time that the signal was shifted. In order to even out differences in auditory saliency between effects, we define upper bounds for the effect amount separately for each effect (see Table \ref{stab:effect_amount}). These bounds were manually adjusted by one of the authors and further tested by all other authors. Further, some effects had to be parametrized with additional parameters which are listed in Table \ref{stab:effect_parametrization}.
\begin{table}[]
    \centering
    \begin{tabular}{|c|c|}
        \hline
    Effect & Upper Bound \\
    \hline
        Pitch &  0.5\\
        Tremolo &  0.4\\
        Synthesis Quality &  1.0\\
        Timeshift & 45ms \\
        Vocoder &  0.35\\
        Flanger &  0.78\\
    \hline
    \end{tabular}
    \caption{Upper boundaries for the effect amounts for each effect. Lower boundaries is always 0.}
    \label{stab:effect_amount}
\end{table}

\begin{table}[]
    \centering
    \begin{tabular}{|c|c|c|}
        \hline
    Effect & Parameter & Value \\
    \hline
        Flanger Type 1 & Delay & 1\\
        Flanger Type 1 & Depth & 10\\
        Flanger Type 1 & Frequency & 5\\
        Flanger Type 2 & Delay & 0\\
        Flanger Type 2 & Depth & 50\\
        Flanger Type 2 & Frequency & 0\\
        Flanger Type 3 & Delay & 20\\
        Flanger Type 3 & Depth & 20\\
        Flanger Type 3 & Frequency & 5\\
        Flanger Type 4 & Delay & 1\\
        Flanger Type 4 & Depth & 10\\
        Flanger Type 4 & Frequency & 25\\
        Flanger Type 5 & Delay & 10\\
        Flanger Type 5 & Depth & 0\\
        Flanger Type 5 & Frequency & 0\\
        Vocoder & Carrier Frequency & 30\\
        Vocoder & Harmonics & 1.0\\

    \hline
    \end{tabular}
    \caption{Fixed parameters of the effects.}
    \label{stab:effect_parametrization}
\end{table}

\subsection{Robot labels}
The list of 260 robot labels is depicted in Table \ref{stab:260_attributes_1}--\ref{stab:260_attributes_4}. These labels were taken from psychological literature about human personality traits \cite{trapnell_extension_1990,mccrae_introduction_1992,gosling_TIPI_2003,rammstedt_bfi-10_2007, hofstee_integration_1992, deRaad_pan-cultural_2014}, from the Godspeed Questionnaires that were created specifically for social robots \cite{bartneck_godspeed_2009}, from the personality dimensions that V\"olkel et al. identified for voice assistants \cite{voelkel_developing_2020}, as well as from the AttrakDiff questionnaire that measures user experience \cite{Hassenzahl_attrakdiff:_2003,hassenzahl_beauty_2004}. The references for each label are listed next to it in the aforementioned tables.

In many cases, the sources did not contain the exact word, but a synonym or an antonym of it. These instances are marked with a ' respectively * symbol next to the source's abbreviation.

Furthermore, several labels were added after a small-scale pilot study for the labeling task showed a tendency for labeling robots with their visible properties. In particular, labels were added for the (apparent) sex, gender presentation and age since they were also expected to relate to the voices. Other examples include "animallike" due to the large number of non-human robots, as well as adjectives referring to the size or to attractiveness in general. In the tables, these are marked with a $\circ$ symbol.

\subsection{Number of robots in previous studies}
\label{sup_number_robots}
The present study incorporates a much larger number of robots (350) than previous research (max. 8 different robots):
\begin{itemize}
    \item Alonso-Martín et al. \cite{alonso2019online}: 3 robots
    \item Alonso-Martín et al. \cite{alonso2020four}: 3 robots
    \item Aylett et al. \cite{aylett2019right}: 3 robots
    \item Häring et al. \cite{haering:et:al:2014}: 1 robot
    \item James et al. \cite{james2020empathetic}: 1 robot
    \item Kuchenbrandt et al. \cite{kuchenbrandt:et:al:2014}: 1 robot
    \item McGinn \& Torre \cite{mcginn2019can}: 8 robots
    \item Powers \& Kiesler \cite{powers2006advisor}: 4 robots
    \item Ritschel et al. \cite{ritschel2019personalized}: 1 robot
    \item Torre et al. \cite{torre2018trust}: 1 robot
\end{itemize}

\subsection{Implicit bias awareness training}
\label{sup_implicit_bias}
We adapted the Implicit Association Test (IAT) \cite{Schnabel2008} to assess implicit biases in our participants. This test measures the association between words (either passive and active attributes or positive and negative words). The task is done on randomly selected six images from each target in a pair. Possible target pairs are adult vs. children, cats vs. dogs, and men vs. women. We focus on the target pair men vs. women, since the implicit biases were pronounced in the collected data. Since some attributes in the test are rarely used – such as ``servile'' or ``obsequious'' –, we select the 10 most frequent words from the 16 active and 16 passive words. The selected words occur at least once per one million. The selected active words are: ``strong'', ``active'', ``effective'', ``mobile'', ``alive'', ``dynamic'', ``animated'', ``lively'', ``potent'', and ``energetic''. The selected passive words are: ``gentle'', ``passive'', ``inactive'', ``tame'', ``compliant'', ``yielding'', ``meek'', ``submissive'', ``obedient'', and ``controllable''.

\begin{figure}[ht]
  \centering
  \includegraphics[width=1\linewidth]{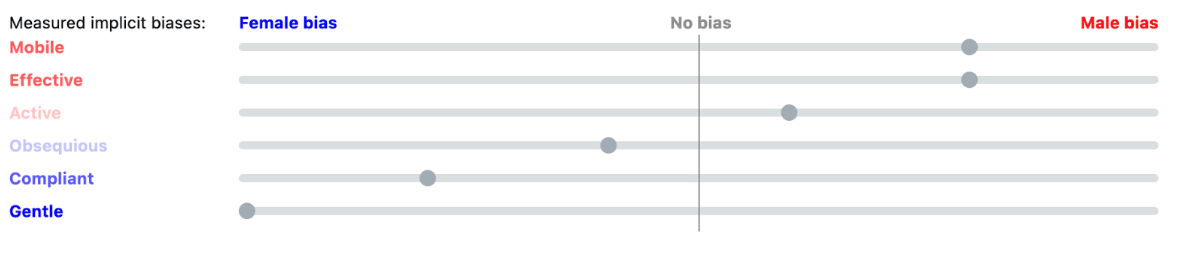}
  \caption{Example of measured biases in Implicit Association Test.}
  \label{sfig:step_bias}
\end{figure}

For each participant, we randomly select three passive and three active words. We tell the participant ``In this task, you will be shown a word and two images. Your task is to choose the image that fits the word better.'' On every page, participants see a random male or female image (from the 10 images per gender in the IAT). The order of the female image (left or right) was random. On top of the image, participants see one of the 6 selected words. The images are shown for 2 seconds and automatically disappear. Participants use the keys on their keyboard to indicate if the left (key A) or the right image (key L) fits best to the attribute. Each of the six attributes is visited 5 times. After the participants complete all 30 trials, we show the measured biases (see Figure \ref{sfig:step_bias} for an example).

\begin{figure}
    \centering
    \includegraphics{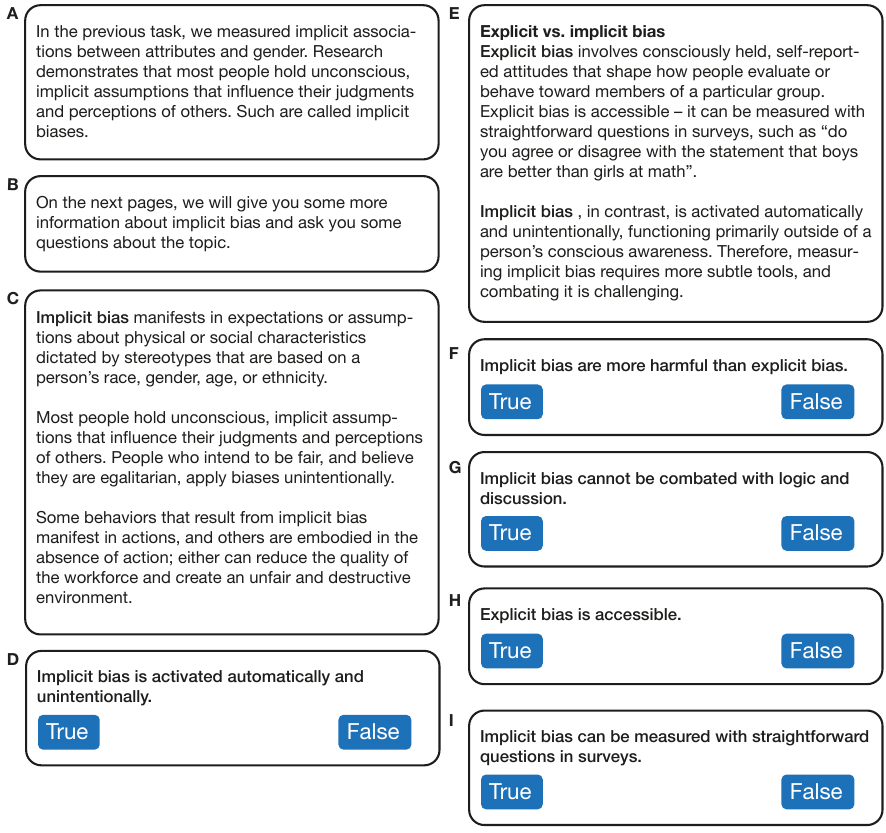}
    \caption{Implicit bias pages part 1. The letters indicate page order.}
    \label{sfig:implicit_bias1}
\end{figure}

\begin{figure}
    \centering
    \includegraphics{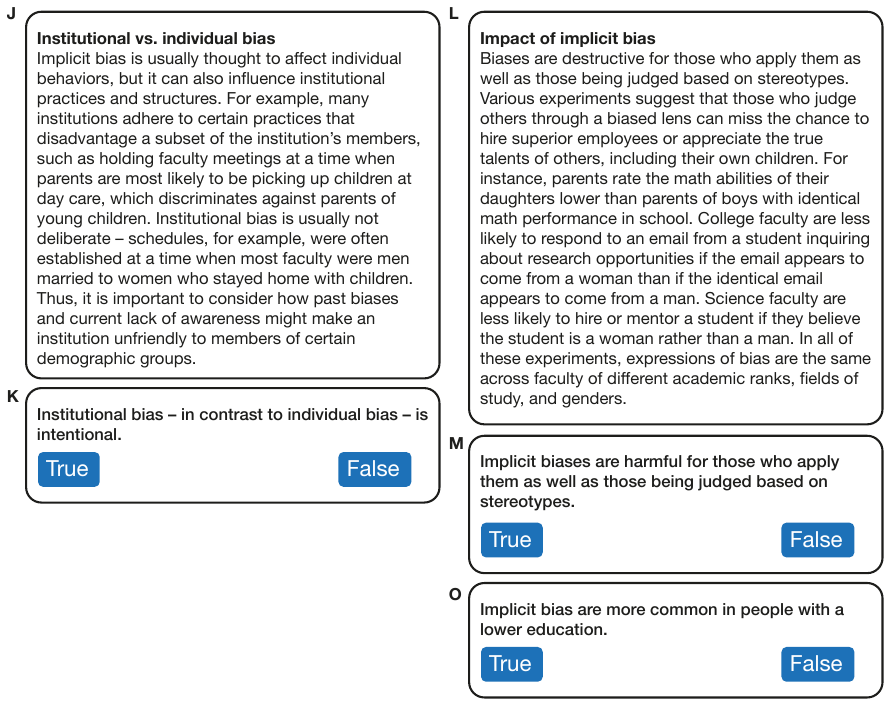}
    \caption{Implicit bias pages part 2. The letters indicate page order.}
    \label{sfig:implicit_bias2}
\end{figure}

Participants proceed with an implicit bias training. For this, we selected the following excerpts from the implicit bias fact sheet from the White House Office of Science and Technology Policy\footnote{\url{https://obamawhitehouse.archives.gov/sites/default/files/microsites/ostp/bias_9-14-15_final.pdf}} (Figure \ref{sfig:implicit_bias1}--\ref{sfig:implicit_bias2})

In both experiments, on average, participants answered 6/8 questions correctly, indicating they carefully read the implicit bias awareness sheet. In both experiments, we did not find a significant difference across the terms after correcting for multiple comparisons (Bonferroni).

\section{Create Voices using Gibbs Sampling with People}
\subsection{Instructions Main Experiment}
\label{sup_gsp_instructions}
\begin{figure}
    \centering
    \includegraphics{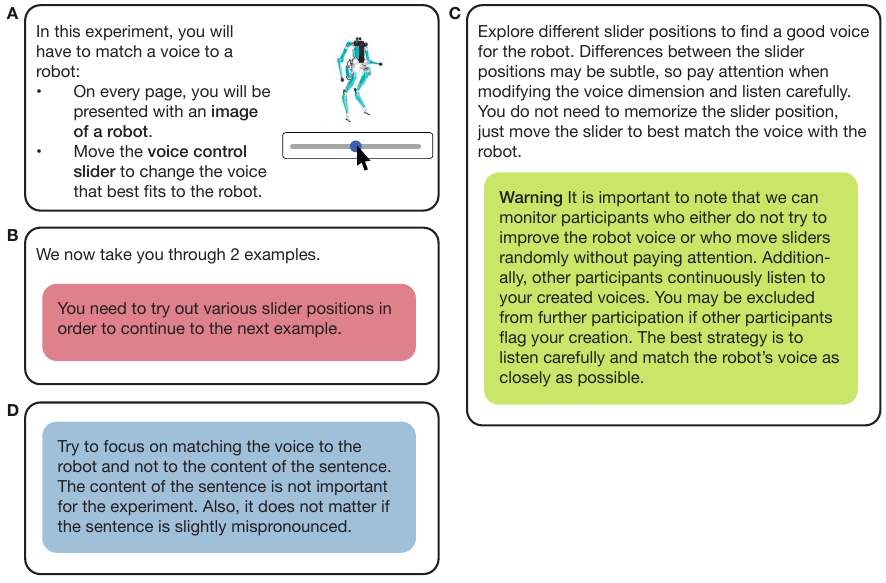}
    \caption{Instructions for GSP experiment.}
    \label{sfig:gsp_instructions}
\end{figure}

The experiments proceeded as follows: Upon completion of the consent form and the pre-screening tasks, participants received instructions regarding the main experiment (see Figure \ref{sfig:gsp_instructions}).

\begin{figure}[ht]
  \centering
  \includegraphics[width=0.8\linewidth]{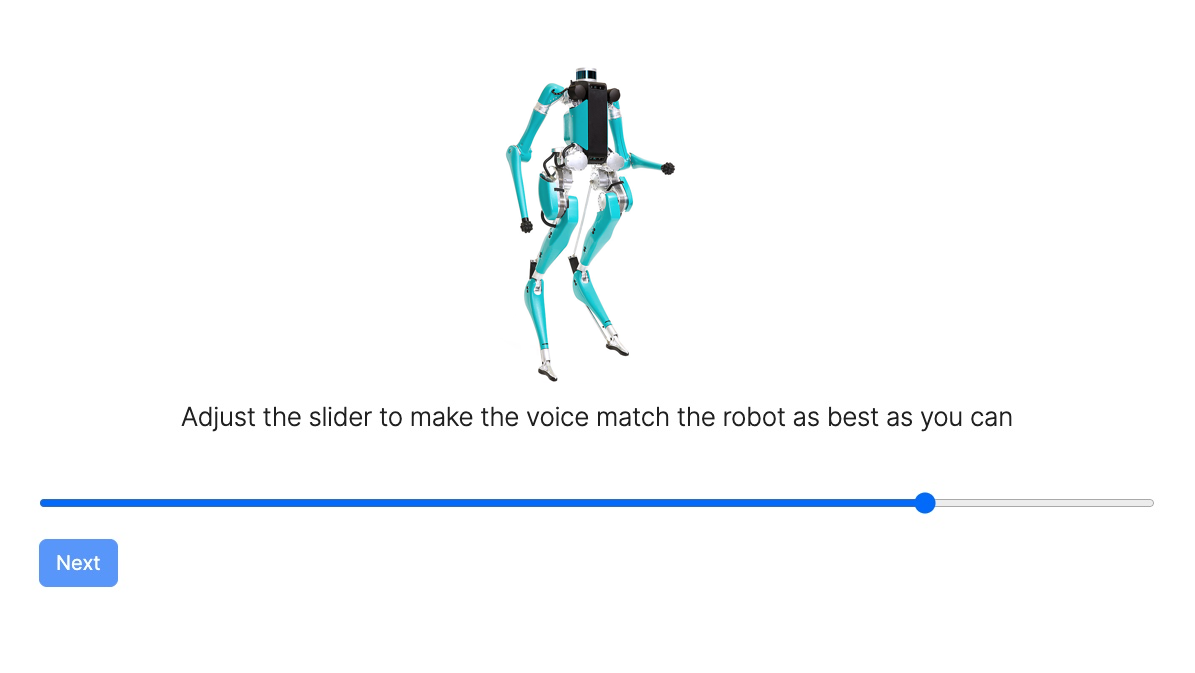}
  \caption{Example trial in the GSP experiment.}
  \label{sfig:gsp_screenshot}
\end{figure}

As described in the instructions, in each trial, participants move a slider corresponding to one voice dimension and have to move the slider to the position such that the obtained voice maximally matches the robot. A screenshot of the task is shown in Figure \ref{sfig:gsp_screenshot}.

\subsection{Different ways to summarize data from previous iterations in GSP experiments}
\label{sup_gsp_aggregation}
Harrison et al. \cite{harrison2020gsp} found that aggregating data from multiple participants in Gibbs Sampling with People (GSP) can improve sampling quality by reducing noise. However, the choice of how to summarize the aggregated data (e.g., mean, median, kernel density estimate) can impact the results. They used mean aggregation for a GSP experiment involving color-matching tasks (Experiment 1 in their paper) but selected the most common value, for face generation experiments (Experiment 4 in their paper), which they deemed more suitable for complex, multi-modal data. While we couldn't directly use their most common value aggregation approach due to a limited number of responses, we used a similar approach of median aggregation, which ensures that only played voice configurations propagate to the next iteration. Median aggregation, like the choice of the most frequent value in Harrison et al.'s generative face domain, is appropriate for our domain, because it prevents the selection of an unpopular intermediate value, as stimulus generation is time-consuming and slider changes may not be smooth. 

%
\subsection{Instructions Validation Experiment}
\label{sup_validation_instructions}

\begin{figure}
    \centering
    \includegraphics{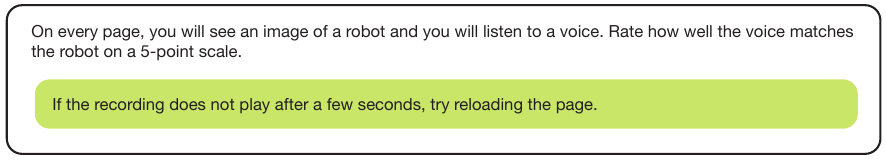}
    \caption{Instructions for validation experiment.}
    \label{sfig:validation_instructions}
\end{figure}

\begin{figure}[ht]
  \centering
  \includegraphics[width=0.8\linewidth]{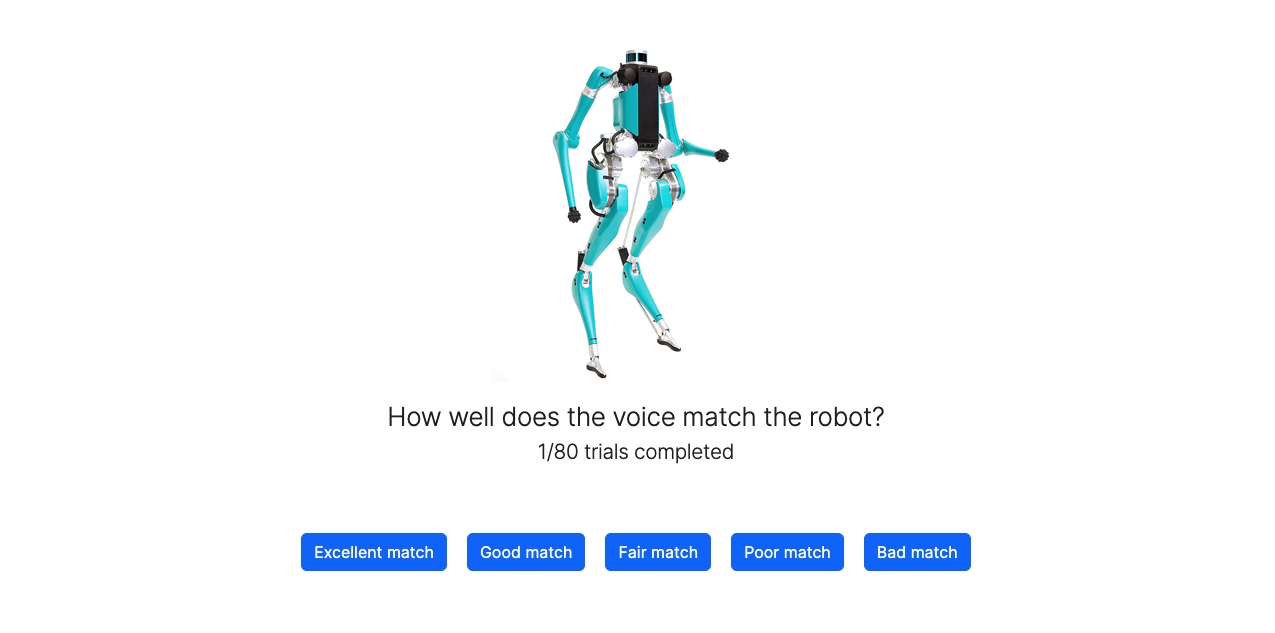}
  \caption{Example trial in the validation experiment.}
  \label{sfig:gsp_validation_screenshot}
\end{figure}

The instructions for the experiment are shown in Figure \ref{sfig:validation_instructions}. A screenshot of the task is shown in Figure \ref{sfig:gsp_validation_screenshot}.

\section{Annotate Dimension using STEP-Tag}
\subsection{Instructions}
\label{sup_step_instructions}
\begin{figure}
    \centering
    \includegraphics{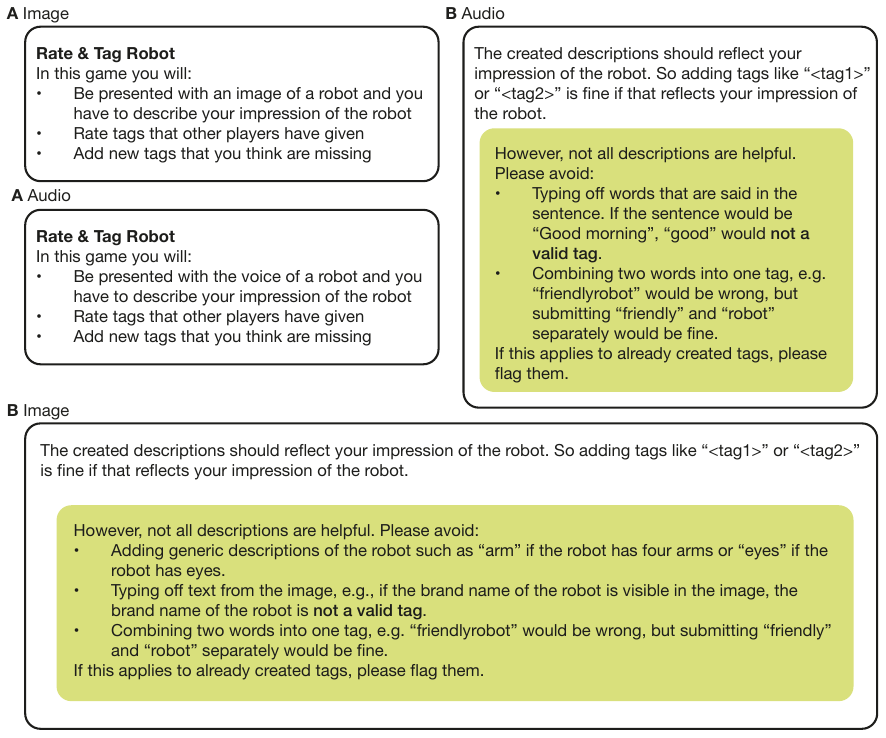}
    \caption{Instructions for STEP experiment part 1/2.}
    \label{sfig:step_instructions1}
\end{figure}

\begin{figure}
    \centering
    \includegraphics{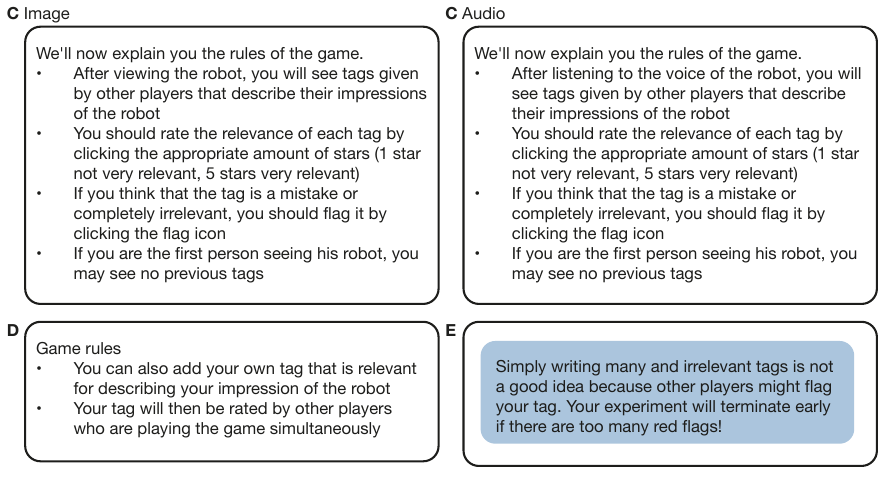}
    \caption{Instructions for STEP experiment part 2/2.}
    \label{sfig:step_instructions2}
\end{figure}

The experiments proceeded as follows: upon completion of the consent form and the pre-screening tasks, participants received instructions regarding the main experiment (Figure \ref{sfig:step_instructions1}--\ref{sfig:step_instructions2}).

``<tag1>'' or ``<tag2>'' are randomly selected from the following terms which were commonly used in a previous pilot:
\begin{itemize}
    \item friendly,
    \item cute,
    \item functional,
    \item weird,
    \item humanlike,
    \item creepy,
    \item strange,
    \item odd,
    \item scary,
    \item unsettling,
    \item uncanny,
    \item powerful
\end{itemize}

\begin{figure}[ht]
  \centering
  \includegraphics[width=0.5\linewidth]{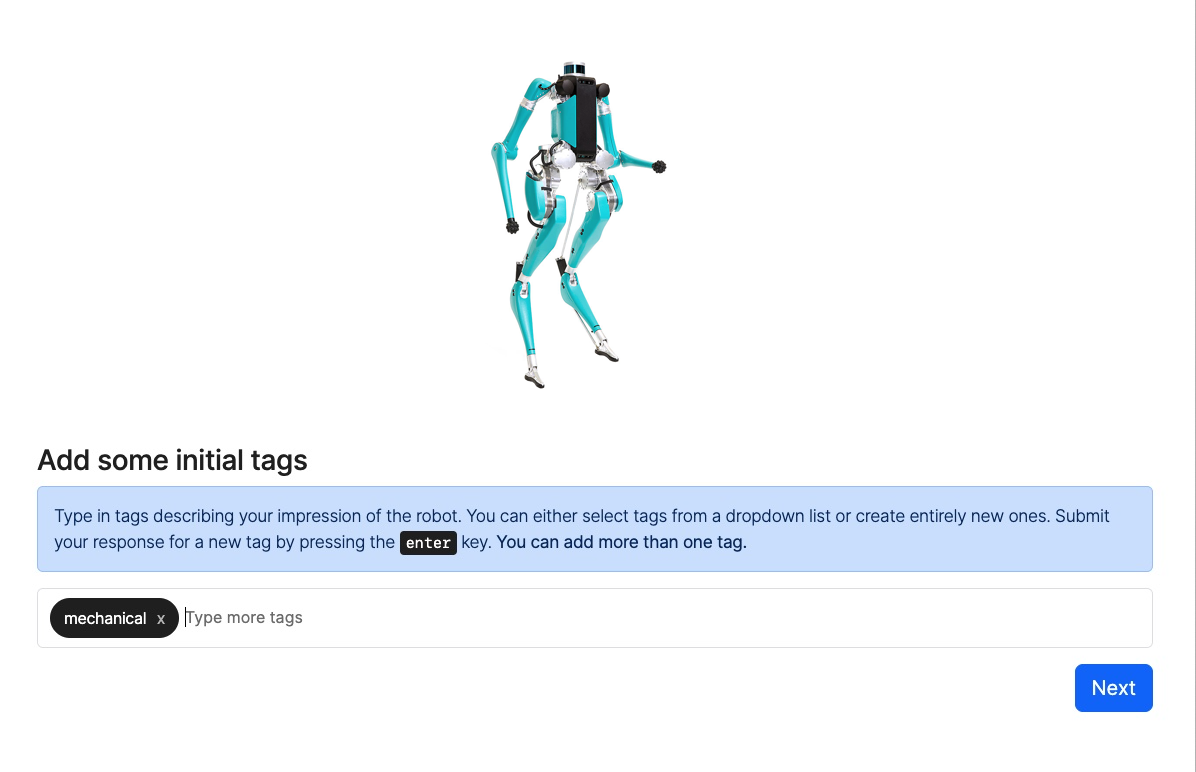}
  \caption{Example trial in the image STEP experiment.}
  \label{sfig:step_screenshot}
\end{figure}

A screenshot of the task is shown in Figure \ref{sfig:step_screenshot}.

\subsection{STEP with implicit bias training}
\label{sup_step_bias}
\begin{figure}[ht!]
  \centering
  \includegraphics[width=\linewidth]{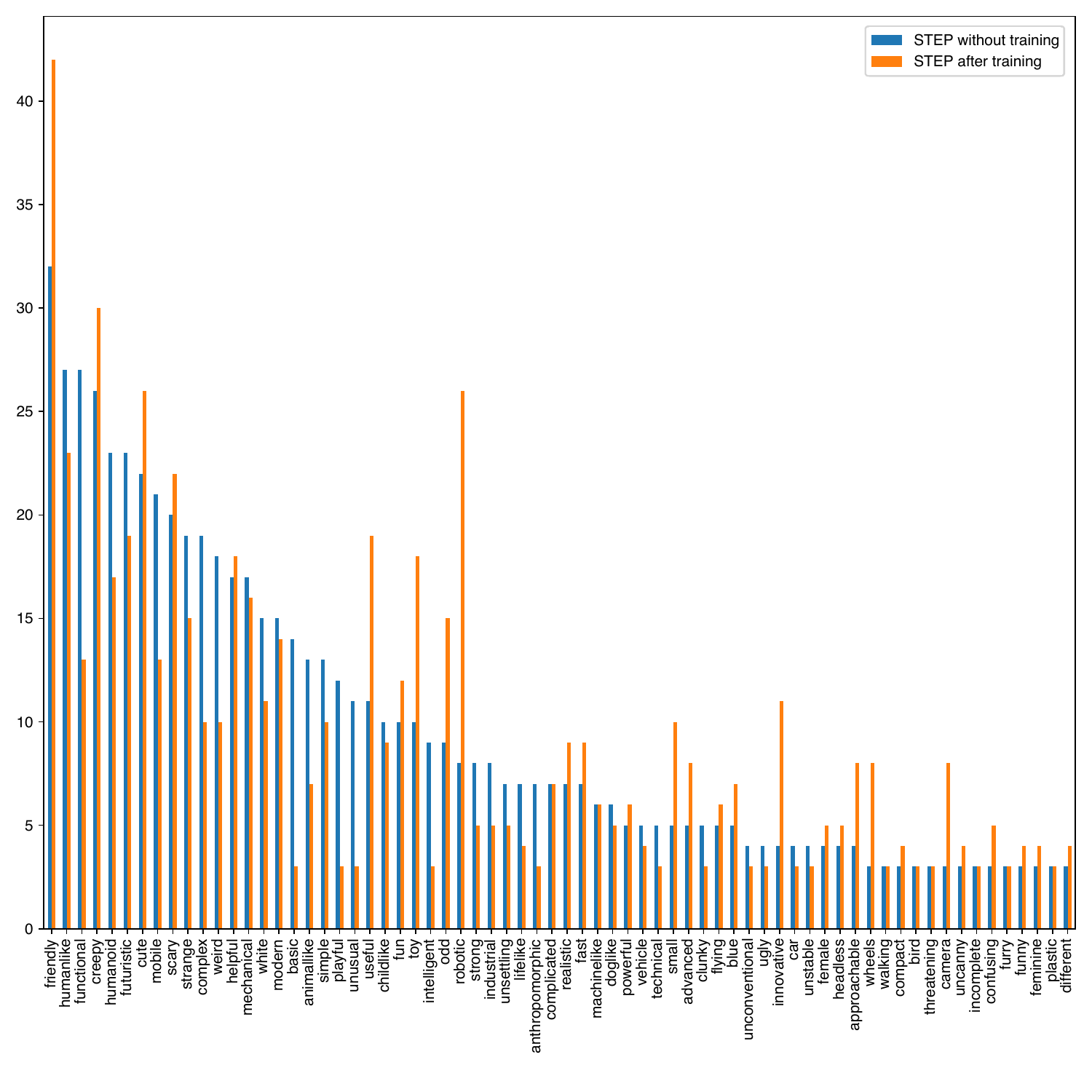}
  \caption{Frequency of terms with or without implicit bias training.}
  \label{sfig:step_frequency}
\end{figure}

There is considerable overlap between the tags obtained in both STEP-Tag experiments. As shown in Figure \ref{sfig:step_frequency}, there is a strong correlation (\textit{r} = .78) between the frequency of the tags. Indicating that the same tags were used to describe the robots. If the data of the STEP experiment after the awareness training had been used to compile the list of 40 terms, it would have led to the replacement of only a single term. The term ``functional'' would have been replaced by ``modern''. The results show that the awareness training had very little effect on the collected data.

\section{Rate Robots}

\subsection{Labels}
\label{sup_used_labels}
The 40 used dimensions and their sources are listed in Supplementary Table \ref{stab:40_dimensions}.

\subsection{Instructions}
\label{sup_dense_instructions}

\begin{figure}
    \centering
    \includegraphics{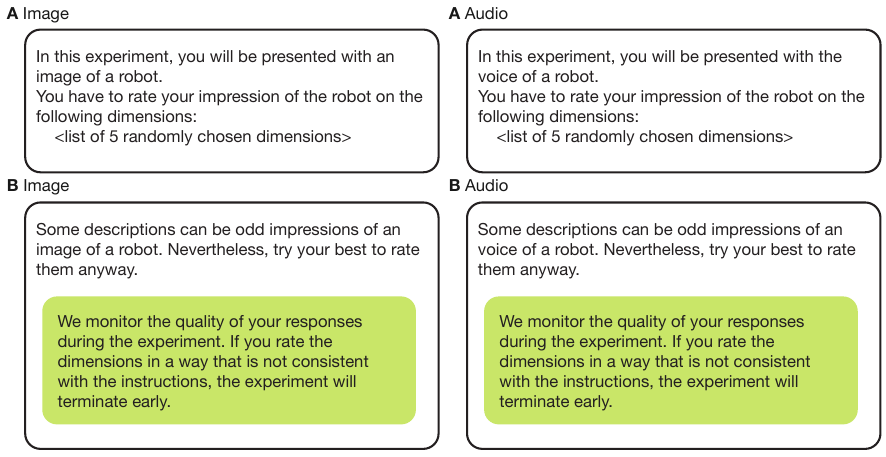}
    \caption{Instructions for dense rating experiment.}
    \label{sfig:dense_instructions}
\end{figure}

\begin{figure}[ht!]
  \centering
  \includegraphics[width=0.5\linewidth]{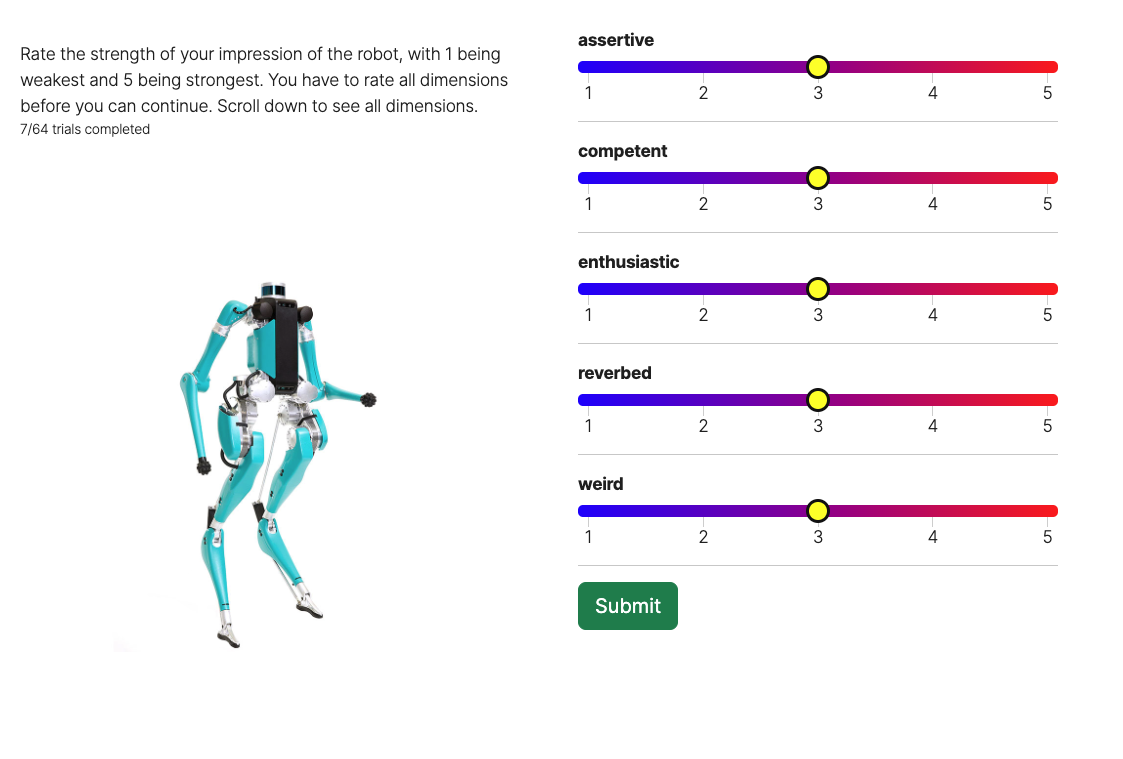}
  \caption{Example trial in the dense rating experiment.}
  \label{sfig:dense_screenshot}
\end{figure}

The instructions are depicted in Figure \ref{sfig:dense_instructions}, a screenshot of the task is shown in Figure \ref{sfig:dense_screenshot}.

\subsection{Consistency between STEP-Tag and dense rating}
\label{sup_step_consistency}
\begin{figure}[h]
    \centering
    \includegraphics{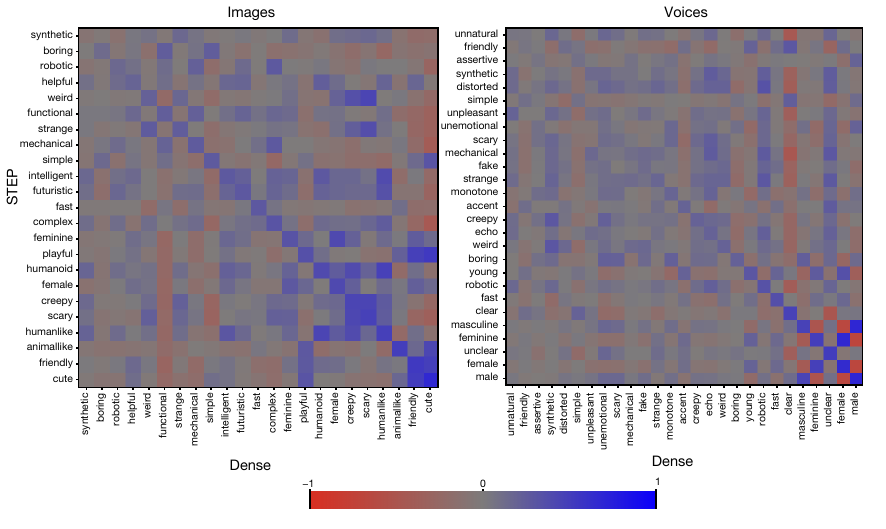}
    \caption{Consistency between STEP-Tag and dense rating. Correlation matrices are sorted by the strength of the diagonal.}
    \label{sfig:STEP-consistency}
\end{figure}

We computed the correlation between the STEP-Tag ratings and the dense ratings. For most dimensions, the value on the main diagonal is relatively large compared with other values, as shown in Figure \ref{sfig:STEP-consistency}. This suggests that dimensions were rated similarly across the two experiments. Furthermore, we found block-like structure for words with a similar meaning (e.g., female and feminine) or words with opposite meanings (e.g., male and female), indicating semantic clusters of terms. Some of these clusters also extend beyond the exact meaning or antonym, , for example ``cute'' (STEP-Tag) is not only highly correlated with ``cute'' (Dense), but also with ``friendly'' and ``playful''. Interestingly, the strength of the diagonal differs across the two modalities. For example, ``scary'' has a rather strong correlation in the visual modality, but is weaker in the voice modality. Generally, the correlation seems strongest for dimensions that are most salient in that modality: For example, the biological sex of a speaker and the clarity of their voice are salient, and features such as ``humanlike'', ``animallike'', or ``friendly'' have clear visual cues.

\subsection{Consistency across modalities}
\begin{figure}
    \centering
    \includegraphics[width=\linewidth]{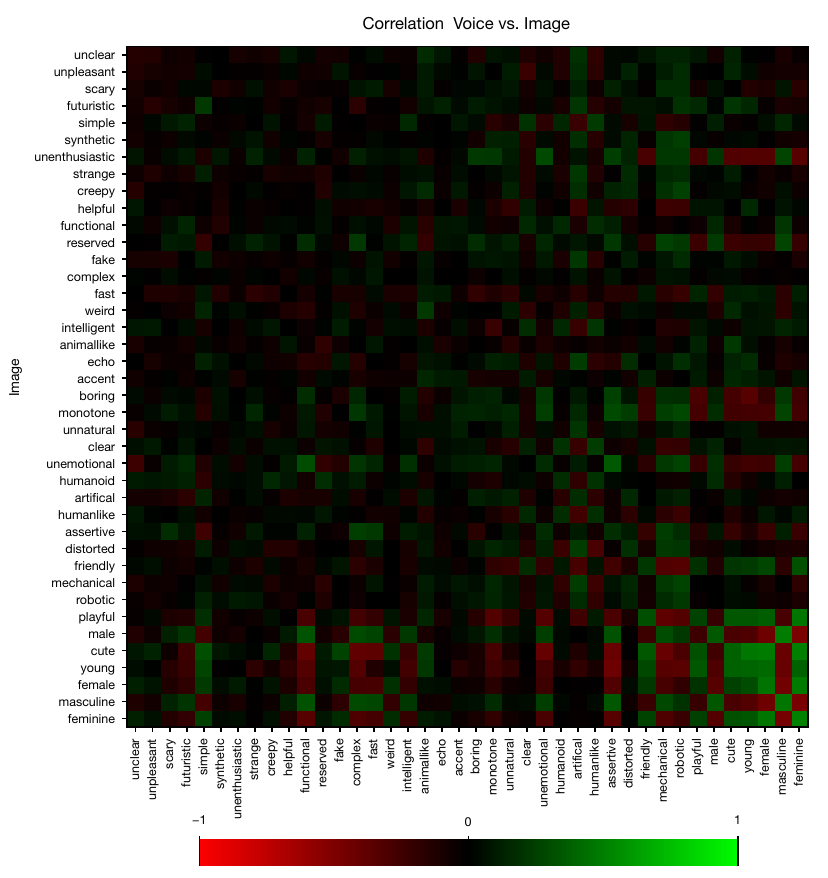}
    \caption{Correlation matrix between the ratings in the image and voice modality. An interactive version of this plot is available at: \url{https://robotvoice.s3.amazonaws.com/compare.html}.}
    \label{sfig:consistency_across_modalities}
\end{figure}

Figure \ref{sfig:consistency_across_modalities} shows the same data as in Figure \ref{fig:dense_results}D but now the dimensions are sorted by the strength of the diagonal.

\label{sup_consistency_across_modalities}

\subsection{Generalizability of the findings}
\label{sup_dense_generalizability}
\begin{figure}[h]
    \centering
    \includegraphics{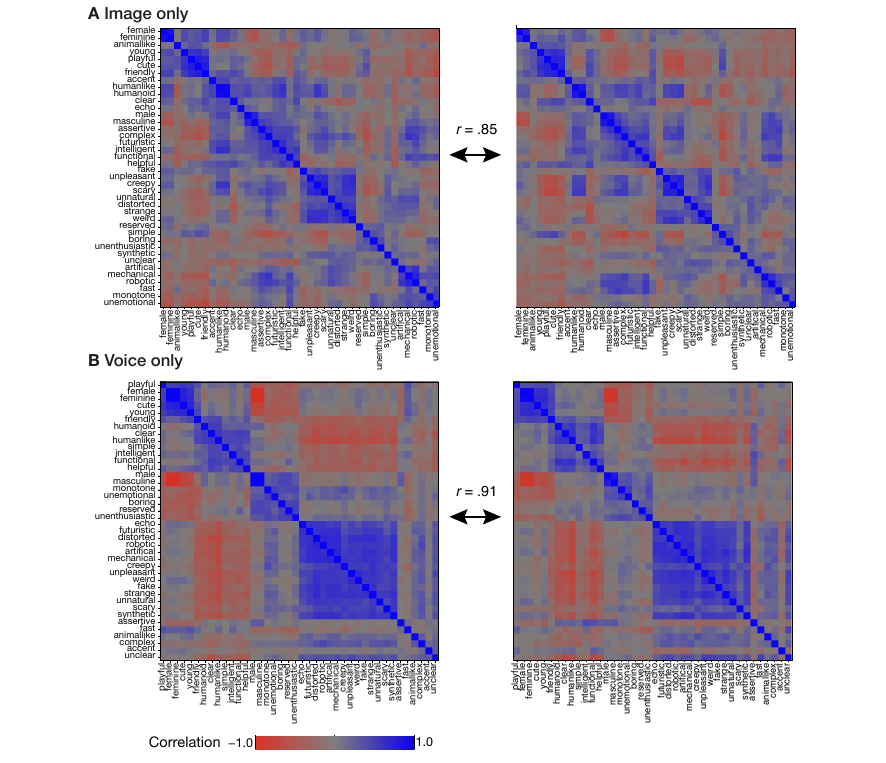}
    \caption{Replication of dense rating experiments on 175 new robot images (\textbf{A}) and on the initial random voices (\textbf{B}).}
    \label{sfig:dense_replication}
\end{figure}

To assess the generalizability of the findings, we run the dense rating on 175 new robot images (see Supplementary Materials \ref{sup_new_robot_dataset}) and on the initial random voices (i.e., iteration 0). We show in Figure \ref{sfig:dense_replication}, that the obtained correlation matrices strongly correlate with the initial correlation matrices: \textit{r} = 0.85 for the image and \textit{r} = 0.91 for the voice modality. These findings indicate that the obtained correlations across the terms are robust across databases.

\subsection{Acoustic correlates}
\label{sup_acoustic_correlates}
\begin{figure}
    \centering
    \includegraphics[width=0.5\linewidth]{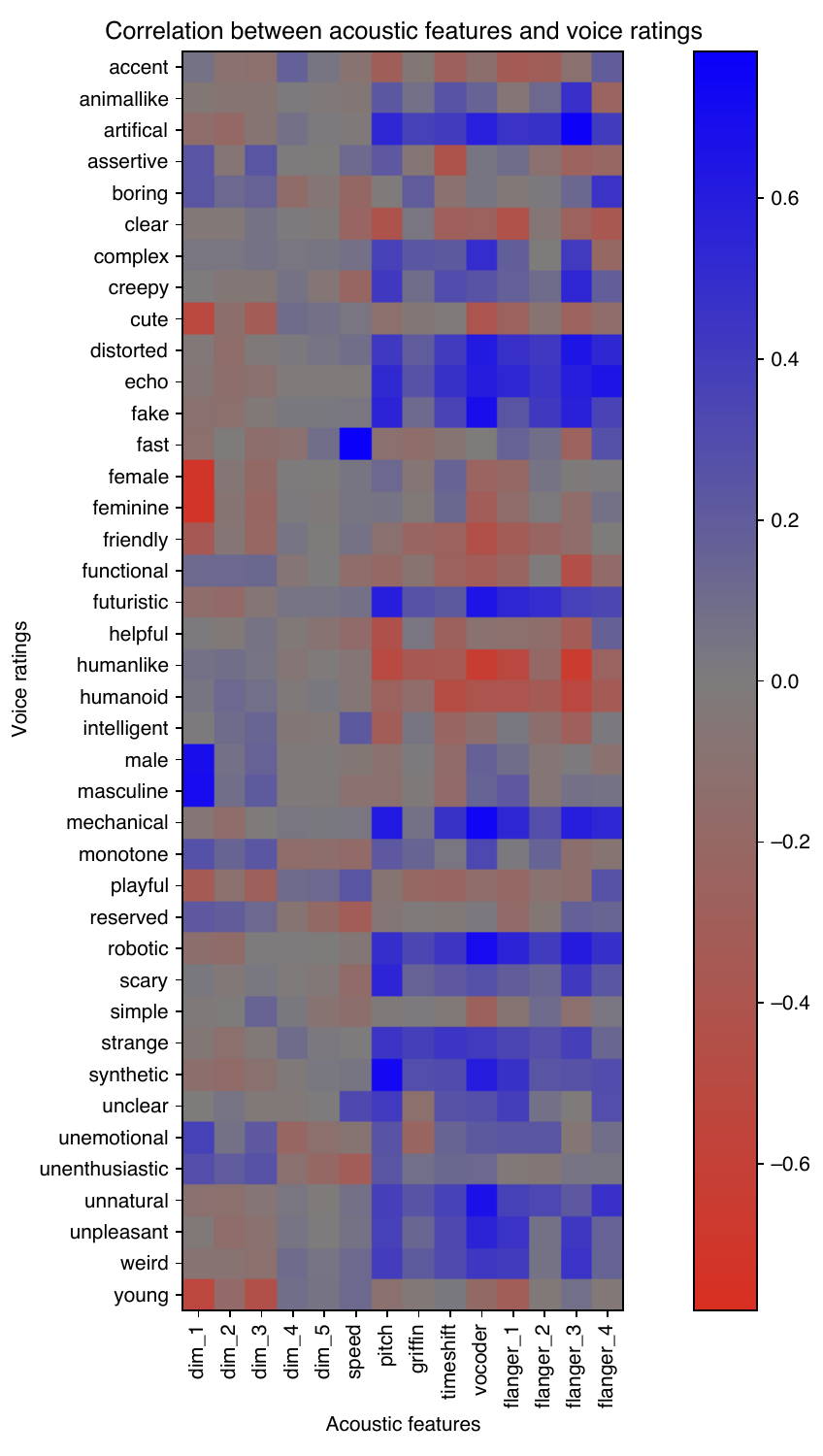}
    \caption{Correlations between voice features and perceptual dimensions.}
    \label{sfig:acoustic_correlates}
\end{figure}

Figure \ref{sfig:acoustic_correlates} shows the correlations between the voice features and the perceptual dimensions. The first and third voice dimensions are correlated with an older male voice. The other latent voice dimensions do not correlate strongly with the 40 dimensions. Speaking speed correlates with ``fast'', ``unclear'', ``playful'', and ``intelligent''. All acoustic effects strongly correlate with ``artificial'', ``robotic'', ``strange'', and ``unnatural''.

\subsection{Dense rating with implicit bias training}
\label{sup_dense_bias}
\begin{figure}[h]
  \centering
  \includegraphics[width=\linewidth]{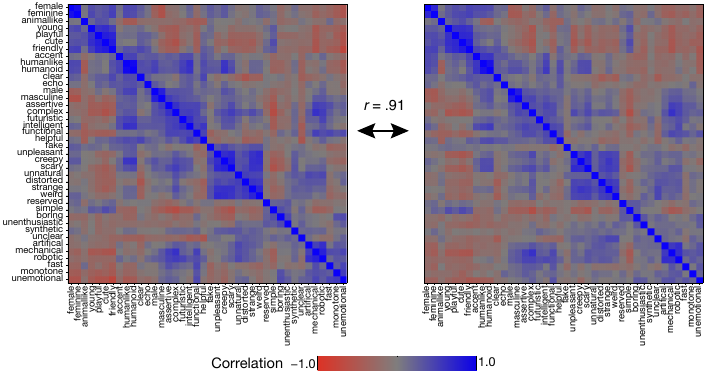}
  \caption{Correlation across correlation matrices in dense image rating experiment with or without implicit bias training.}
  \label{sfig:bias_dense}
\end{figure}

As shown in Figure \ref{sfig:bias_dense}, the implicit bias awareness training barely changed the correlations across terms as indicated by the high correlation across the upper triangles without diagonals (\textit{r} = .91). While participants closely followed the implicit bias training (correct comprehension questions), it barely influenced the correlations across the terms.

\subsection{Replace dense rating by deep learning model}
\label{sup_dense_clip}

To investigate if we can replace the dense rating procedure by a deep learning model, we provide two analyses on the pretrained CLIP model.\footnote{\url{https://github.com/openai/CLIP}}

In the first analysis, we compute the correlation between the cosine similarity computed on the dense image rating (which was used to predict a voice) and the cosine similarity of the image embedding. For both the old (\textit{r} = .58) and the new set of 175 images (\textit{r} = .51) we found a moderate correlation between the upper triangles of both cosine similarity matrices. This indicates that CLIP provides a fair proxy for the perceived similarity of robots.

\begin{figure}[h]
    \centering
    \includegraphics{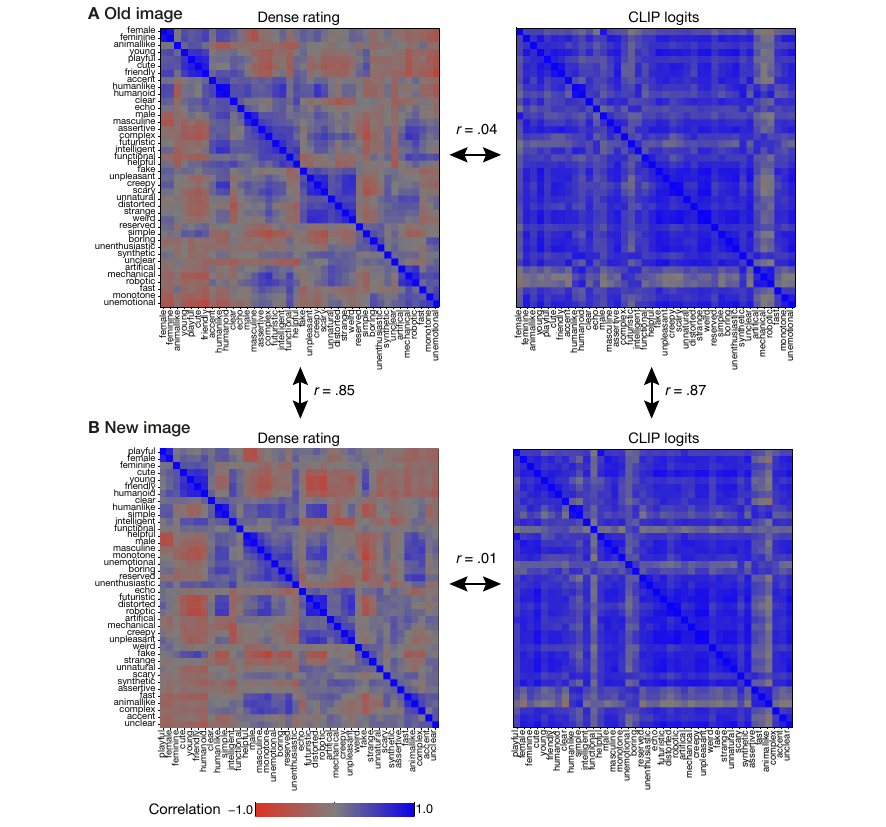}
    \caption{Correlation across 40 dimensions for human dense labeling and CLIP.}
    \label{sfig:sfig_dense_clip}
\end{figure}

In the second analysis, we use CLIP to do the dense rating. For each image, we obtain a logit value for each of the 40 dimensions. In Figure \ref{sfig:sfig_dense_clip} we show the correlations across the 40 dimensions. Generally, the correlations across the dimensions are high in CLIP. There is a similarly strong correlation across the CLIP results across datasets (\textit{r} = .85) compared to the dense rating results (\textit{r} = .87). The results show that the correlational structure across the terms is consistent across the datasets, but varies greatly between CLIP and the human dense rating.

\section{Prediction}
\subsection{Instructions}
\label{sup_prediction_instructions}
The instructions are identical to the GSP validation experiment (Section \ref{sup_validation_instructions}).

\subsection{Prediction per participant}
\label{sup_prediction_by_participant}
\begin{figure}[h]
    \centering
    \includegraphics[width=\linewidth]{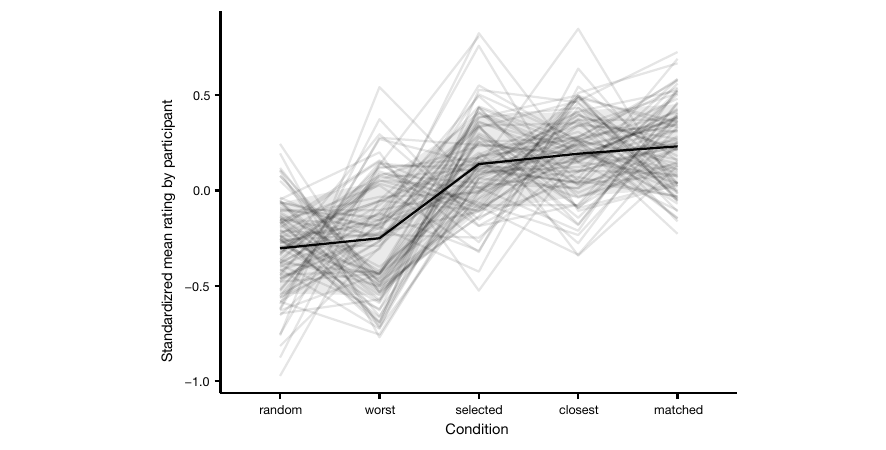}
    \caption{Prediction per participant. Mean standardized rating by participants and condition. Single lines depict single participants. The dark black line is the average across participants. The shaded area is the standard deviation across participants.}
    \label{sfig:prediction_by_participant}
\end{figure}

To assess if the prediction result can also be found in single participants, we z-scored all ratings by each participant. We then computed the mean per participant and condition which are depicted as thin lines in Figure \ref{sfig:prediction_by_participant}. As shown in Figure \ref{sfig:prediction_by_participant}, most participants show the trend consistent with overall mean (thick line).

\subsection{Factor analysis}
\label{sup_prediction_factor_analysis}
To predict the closest robot based on perceptual dimensions, we performed a factor analysis on the 40-dimensional image ratings. Of the 40 dimensions, 39 are correlated at least 0.3 with at least one other feature, suggesting reasonable factorability. The Kaiser–Meyer–Olkin measure of sampling adequacy is 0.85, and Bartlett’s test of sphericity is significant (5013.3, $p$ < 0.001). Therefore, we applied factor analysis with Varimax (orthogonal) rotation.

\begin{figure}[h]
    \centering
    \includegraphics{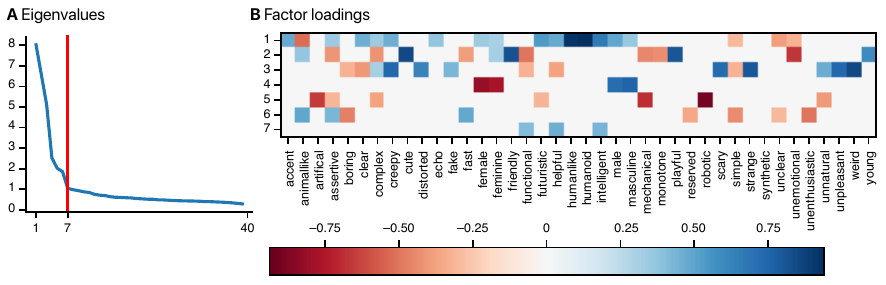}
    \caption{Factor analysis. \textbf{A} Eigenvalues plot. \textbf{B} Factor loading plot. Weak loadings (< 0.3) are omitted for the readability of the figure.}
    \label{sfig:factor_analysis}
\end{figure}

We selected a seven-factor solution because the first seven eigenvalues are > 1 (Figure \ref{sfig:factor_analysis}A). The factors explain 21\%, 21\%, 20\%, 10\%, 10\%, 7\%, and 2\% of the variance (91\% in total). Factor 1, ``humanlike'', mainly loads on humanlike and humanoid (see Figure \ref{sfig:factor_analysis}B for the loading plot, the factor name is given by the dimension with the strongest loading). Factor 2, ``cute'', loads mainly on cute, friendly, playful, and young. Factor 3, ``creepy'', loads on creepy, distorted, scary, strange, unpleasant, and weird. Factor 4, ``gender'', positively loads on male and negatively on female. Factor 5, ``natural'' negatively loads on artificial, mechanical, and robotic. Factor 6, ``fast'', mildly loads on animallike, assertive, and fast. Factor 7, ``functional'', mildly loads on functional, helpful, and intelligent.

\begin{table}[htb]
    \begin{tabular}{ll|cc}
        \textbf{source(s)} & \textbf{adjective} & \multicolumn{2}{c}{\textbf{in STEP}} \\
        \hline \hline							
        GS, VA	&	friendly	&	$\square$	&	$\triangledown$	\\
        GS, VA'	&	mechanical	&	$\square$	&	$\triangledown$	\\
        AD, TIPI, VA'	&	simple	&	$\square$	&	$\triangledown$	\\
        \hline							
        TIPI, BFI-10	&	reserved	&		&	$\triangledown$	\\
        IASR-B5, FFM, VA	&	assertive	&		&	$\triangledown$	\\
        GS, VA	&	fake	&		&	$\triangledown$	\\
        GS	&	unpleasant	&		&	$\triangledown$	\\
        GS	&	artifical	&		&	$\triangledown$	\\
        VA	&	synthetic	&		&	$\triangledown$	\\
        PCPS*, HRG	&	unemotional	&		&	$\triangledown$	\\
        AD	&	clear	&		&	$\triangledown$	\\
        TIPI*	&	unenthusiastic	&		&	$\triangledown$	\\
        AD'	&	boring	&		&	$\triangledown$	\\
        GS*	&	unnatural	&		&	$\triangledown$	\\
        AD*	&	unclear	&		&	$\triangledown$	\\
        $\circ$	&	masculine	&		&	$\triangledown$	\\
        $\circ$	&	male	&		&	$\triangledown$	\\
        $\circ$	&	young	&		&	$\triangledown$	\\
        $\circ$	&	feminine	&		&	$\triangledown$	\\
        $\circ$	&	female	&		&	$\triangledown$	\\
        \hline							
        VA	&	playful	&	$\square$	&		\\
        GS, VA	&	intelligent	&	$\square$	&		\\
        VA	&	helpful	&	$\square$	&		\\
        GS	&	humanlike	&	$\square$	&		\\
        TIPI	&	complex	&	$\square$	&		\\
        $\circ$	&	animallike	&	$\square$	&		\\
        
    \end{tabular}
    \begin{tabular}{llp{2cm}lll}    \\
         \multicolumn{2}{l}{\textbf{markers:}} &&  \textbf{sources: }  &                                   \\
         \\
         $\square$ & in STEP-images        &&  IASR-B5 \cite{trapnell_extension_1990}
                                                & HRG \cite{hofstee_integration_1992}
                                                 &   AD \cite{Hassenzahl_attrakdiff:_2003}\\
         $\triangledown$ & in STEP-voices  &&  FFM \cite{mccrae_introduction_1992}
                                                & PCPS \cite{deRaad_pan-cultural_2014}
                                                 &   AD-BG \cite{hassenzahl_beauty_2004}\\
                         &                &&  TIPI \cite{gosling_TIPI_2003}
                                                 &   GS \cite{bartneck_godspeed_2009} &\\
         *       & antonym in source     &&   BFI-10 \cite{rammstedt_bfi-10_2007} 
                                                &   VA \cite{voelkel_developing_2020} &\\        
         '       & synonym in source     &&   & \\
         $\circ$ & added after pilot study  &&   &\\
         \\
     \end{tabular}
    \caption{List of attributes compiled from literature and pilot study. Part 1/4, showing those of the original 260 attributes that were also used by participants in the labeling task.}
    \label{stab:260_attributes_1}
\end{table}

\begin{table}[htb]
    \begin{minipage}[t]{0.4\textwidth}
    \begin{tabular}{ll}
        \textbf{source(s)} & \textbf{adjective} \\
        \hline \hline							
        TIPI	&	uncreative	\\
        AD	&	conservative	\\
        TIPI	&	open to experience	\\
        IASR-B5	&	unphilosophical	\\
        IASR-B5	&	unreflective	\\
        AD, TIPI	&	conventional	\\
        FFM*, AD, BFI-10*	&	unimaginative	\\
        AD, TIPI*, VA	&	creative	\\
        IASR-B5*, BFI-10*, VA	&	artistic	\\
        IASR-B5, FFM, BFI-10'	&	imaginative	\\
        IASR-B5, TIPI*	&	unconventional	\\
        IASR-B5, FFM*	&	unreliable	\\
        AD	&	innovative	\\
        IASR-B5	&	questioning	\\
        IASR-B5	&	philosophical	\\
        IASR-B5	&	reflective	\\
        FFM	&	curious	\\
        AD	&	original	\\
        VA	&	joyful	\\
        IASR-B5	&	broad-minded	\\
        BFI-10, VA	&	lazy	\\
        VA	&	principled	\\
        VA	&	reckless	\\
        AD	&	predictable	\\
        GS	&	irresponsible	\\
        TIPI	&	careless	\\
        IASR-B5, FFM*, TIPI, VA*	&	disorganized	\\
        IASR-B5, FFM*	&	inefficient	\\
        IASR-B5	&	unsystematic	\\
        IASR-B5*, FFM*, VA	&	superficial	\\
        IASR-B5	&	undisciplined	\\
        VA	&	messy	\\
        BFI-10*	&	diligent	\\
        IASR-B5, FFM	&	reliable	\\
        AD	&	unpredictable	\\
        GS	&	responsible	\\
        IASR-B5, FFM, TIPI*, VA	&	organized	\\
        IASR-B5	&	orderly	\\
        IASR-B5, FFM	&	efficient	\\
    \end{tabular}
    \end{minipage}
    \hfill
   \begin{minipage}[t]{0.5\textwidth}
    \begin{tabular}{ll}
        \textbf{source(s)} & \textbf{adjective} \\
        \hline \hline							
        IASR-B5	&	systematic	\\
        IASR-B5, FFM, BFI-10', VA	&	thorough	\\
        IASR-B5, VA*	&	tidy	\\
        TIPI	&	extroverted	\\
        TIPI	&	quiet	\\
        IASR-B5	&	timid	\\
        IASR-B5, VA*	&	forceless	\\
        IASR-B5	&	meek	\\
        FFM	&	talkative	\\
        FFM', VA	&	expressive	\\
        FFM	&	active	\\
        IASR-B5	&	dominant	\\
        VA	&	powerful	\\
        FFM, BFI-10, TIPI*	&	outgoing	\\
        IASR-B5, VA	&	forceful	\\
        IASR-B5	&	firm	\\
        FFM	&	energetic	\\
        FFM, VA	&	enthusiastic	\\
        VA	&	agreeable	\\
        FFM*, BFI-10*, VA	&	distrustful	\\
        VA	&	detached	\\
        GS, VA*	&	unfriendly	\\
        IASR-B5	&	uncharitable	\\
        IASR-B5	&	soft-hearted	\\
        IASR-B5*, FFM*, GS, VA*	&	unkind	\\
        IASR-B5, VA	&	cruel	\\
        FFM*, VA	&	stingy	\\
        IASR-B5	&	ruthless	\\
        GS	&	pleasant	\\
        TIPI*, VA	&	peaceful	\\
        FFM, BFI-10, VA'	&	trusting	\\
        VA	&	benevolent	\\
        VA	&	affectionate	\\
        VA	&	respectful	\\
        IASR-B5, FFM, TIPI	&	sympathetic	\\
        IASR-B5	&	charitable	\\
        IASR-B5	&	iron-hearted	\\
        IASR-B5*, FFM, TIPI, VA	&	warm	\\
        IASR-B5, FFM, GS, VA	&	kind	\\
    \end{tabular}
    \end{minipage}
    \begin{tabular}{llp{2cm}ll}    \\
         \multicolumn{2}{l}{\textbf{markers:}} &&  \textbf{sources: }  &                                   \\
         * & antonym in source        &&  IASR-B5 \cite{trapnell_extension_1990}
                                                 &   AD \cite{Hassenzahl_attrakdiff:_2003} \\
         ' & synonym in source  &&  FFM \cite{mccrae_introduction_1992}
                                                 &   GS \cite{bartneck_godspeed_2009} \\
        $\circ$ & added after pilot study               &&  TIPI \cite{gosling_TIPI_2003}
                                                 &   VA \cite{voelkel_developing_2020}\\
          &      &&   BFI-10 \cite{rammstedt_bfi-10_2007} &       \\
          \\
     \end{tabular}
    \caption{List of attributes compiled from literature and pilot study. Part 2/4, showing a subset of the original 260 attributes that was not used by participants in the labeling task.}
    \label{stab:260_attributes_2}
\end{table}

\begin{table}[htb]
   \begin{minipage}[t]{0.4\textwidth}
    \begin{tabular}{ll}
        \textbf{source(s)} & \textbf{adjective}\\
        \hline \hline
        IASR-B5	&	tender	\\
        FFM	&	appreciative	\\
        FFM	&	forgiving	\\
        FFM	&	generous	\\
        BFI-10	&	sociable	\\
        IASR-B5*, FFM, TIPI*	&	unstable	\\
        IASR-B5, FFM*, TIPI	&	stable	\\
        IASR-B5, BFI-10, VA	&	nervous	\\
        VA	&	temperamental	\\
        FFM	&	impulsive	\\
        IASR-B5, FFM	&	worrying	\\
        IASR-B5, FFM	&	tense	\\
        IASR-B5, FFM*, TIPI*, VA*	&	unanxious	\\
        IASR-B5', VA	&	excitable	\\
        FFM	&	thin-skinned	\\
        IASR-B5*, VA	&	moody	\\
        FFM	&	touchy	\\
        IASR-B5, TIPI, VA	&	calm	\\
        VA	&	stoic	\\
        FFM*, VA	&	deliberate	\\
        IASR-B5, FFM*	&	unworrying	\\
        IASR-B5, BFI-10, VA	&	relaxed	\\
        IASR-B5, FFM, TIPI, VA	&	anxious	\\
        GS	&	incompetent	\\
        GS	&	competent	\\
        GS, VA	&	ignorant	\\
        VA	&	dumb	\\
        GS	&	knowledgeable	\\
        VA	&	useful	\\
        GS	&	natural	\\
        GS	&	machinelike	\\
        GS	&	unconscious	\\
        GS, VA	&	dead	\\
        GS	&	stagnant	\\
        GS	&	inert	\\
        GS	&	conscious	\\
        GS, VA*	&	alive	\\
        GS	&	lively	\\
        GS	&	organic	\\
        GS, VA	&	interactive	\\
    \end{tabular}
    \end{minipage}
    \hfill
    \begin{minipage}[t]{0.5\textwidth}
    \begin{tabular}{ll}
        \textbf{source(s)} & \textbf{adjective} \\
        \hline \hline
        GS, VA	&	responsive	\\
        AD	&	not presentable	\\
        AD	&	unstylish	\\
        AD	&	confusing	\\
        AD	&	cumbersome	\\
        AD	&	complicated	\\
        VA	&	soothing	\\
        AD	&	presentable	\\
        AD	&	valuable	\\
        AD	&	stylish	\\
        AD	&	direct	\\
        AD	&	engaging	\\
        GS	&	moving rigidly	\\
        GS	&	lifelike	\\
        GS	&	moving elegantly	\\
        VA	&	flexible	\\
        GS	&	apathetic	\\
        GS, VA*	&	unintelligent	\\
        GS	&	foolish	\\
        GS	&	sensible	\\
        GS	&	dislike	\\
        GS	&	awful	\\
        GS	&	like	\\
        GS	&	nice	\\
        BFI-10', VA	&	fault-finding	\\
        FFM*, TIPI	&	critical	\\
        TIPI	&	quarrelsome	\\
        FFM, TIPI, VA	&	dependable	\\
        IASR-B5, FFM, TIPI, VA	&	self-disciplined	\\
        GS	&	agitated	\\
        GS	&	surprised	\\
        GS	&	quiescent	\\
        PCPS', HRG	&	thoughtful	\\
        PCPS	&	inattentive	\\
        HRG	&	cautious  \\
        HRG	&	reasonable	\\
        PCPS	&	honest	\\
        HRG	&	weak	\\
        PCPS	&	arrogant	\\
        HRG	&	uncooperative	\\
    \end{tabular}
    \end{minipage}
    \begin{tabular}{llp{2cm}ll}    \\
         \multicolumn{2}{l}{\textbf{markers:}} &&  \textbf{sources: }  &                                   \\
         * & antonym in source        &&  IASR-B5 \cite{trapnell_extension_1990}
                                                 &   AD \cite{Hassenzahl_attrakdiff:_2003} \\
         ' & synonym in source  &&  FFM \cite{mccrae_introduction_1992}
                                                 &   GS \cite{bartneck_godspeed_2009} \\
        $\circ$ & added after pilot study               &&  TIPI \cite{gosling_TIPI_2003}
                                                 &   VA \cite{voelkel_developing_2020}\\
          &      &&   BFI-10 \cite{rammstedt_bfi-10_2007} &       \\ 
          \\
     \end{tabular}
    \caption{List of attributes compiled from literature and pilot study. Part 3/4, showing a subset of the original 260 attributes that was not used by participants in the labeling task.}
    \label{stab:260_attributes_3}
\end{table}

\begin{table}[htb]
    \begin{minipage}[t]{0.4\textwidth}
    \begin{tabular}{ll}
        \textbf{source(s)} & \textbf{adjective} \\
        \hline \hline
        HRG	&	impolite	\\
        HRG	&	cooperative	\\
        HRG	&	polite	\\
        PCPS	&	merciful	\\
        PCPS	&	emotional	\\
        AD-BG	&	ugly	\\
        AD-BG	&	beautiful	\\
        TIPI*, VA*	&	closed-minded	\\
        BFI-10', VA*	&	unartistic	\\
        TIPI'	&	open-minded	\\
        VA*	&	shallow	\\
        IASR-B5*	&	unquestioning	\\
        IASR-B5*	&	narrow-minded	\\
        FFM*	&	uncurious	\\
        FFM*	&	unoriginal	\\
        VA*	&	serious	\\
        VA*	&	unprincipled	\\
        IASR-B5, AD	&	impractical	\\
        IASR-B5'	&	disorderly	\\
        TIPI*	&	careful	\\
        IASR-B5*, AD	&	practical	\\
        IASR-B5', FFM', VA'	&	disciplined	\\
        TIPI*	&	introverted	\\
        FFM*, VA*	&	expressionless	\\
        FFM*	&	passive	\\
        IASR-B5*, FFM*, VA*	&	non-assertive	\\
        IASR-B5*	&	submissive	\\
        VA*	&	powerless	\\
        FFM*	&	non-energetic	\\
        IASR-B5*	&	bold	\\
        VA*	&	disagreeable	\\
        VA*	&	belligerent	\\
        VA*	&	malevolent	\\
        VA*	&	disrespectful	\\
        IASR-B5*, FFM*, TIPI*	&	unsympathetic	\\
        IASR-B5', TIPI*	&	cold	\\
        FFM*	&	unappreciative	\\
        FFM*	&	unforgiving	\\
    \end{tabular}
    \end{minipage}
    \hfill
    \begin{minipage}[t]{0.5\textwidth}
    \begin{tabular}{ll}
        \textbf{source(s)} & \textbf{adjective} \\
        \hline \hline
        IASR-B5*, VA*	&	non-excitable	\\
        FFM*	&	thick-skinned	\\
        VA*	&	unhelpful	\\
        VA*	&	useless	\\
        GS', VA*	&	unresponsive	\\
        GS'	&	agitating	\\
        AD'	&	worthless	\\
        GS'	&	calming	\\
        FFM*, BFI-10*	&	reclusive	\\
        BFI-10*	&	unsociable	\\
        FFM'	&	uncritical	\\
        FFM*, TIPI*, VA*	&	undependable	\\
        TIPI'	&	upset	\\
        TIPI'	&	loud	\\
        HRG*	&	unreasonable	\\
        PCPS*	&	dishonest	\\
        PCPS*	&	attentive	\\
        PCPS*	&	merciless	\\
        GS'	&	intimidating	\\
        GS'	&	upsetting	\\
        GS'	&	reassuring	\\
        $\circ$	&	small	\\
        $\circ$	&	tiny	\\
        $\circ$	&	old	\\
        $\circ$	&	big	\\
        $\circ$	&	tall	\\
        $\circ$	&	distant	\\
        $\circ$	&	involved	\\
        $\circ$	&	changing	\\
        $\circ$	&	constant	\\
        $\circ$	&	repulsive	\\
        $\circ$	&	unattractive	\\
        $\circ$	&	attractive	\\
        $\circ$	&	inelegant	\\
        $\circ$	&	uninteresting	\\
        $\circ$	&	elegant	\\
        $\circ$	&	interesting	\\
        $\circ$	&	uncomfortable	\\
    \end{tabular}
    \end{minipage}
    \begin{tabular}{llp{2cm}ll}    \\
         \multicolumn{2}{l}{\textbf{markers:}} &&  \textbf{sources: }  &                                   \\
         * & antonym in source        &&  IASR-B5 \cite{trapnell_extension_1990}
                                                 &   AD \cite{Hassenzahl_attrakdiff:_2003} \\
         ' & synonym in source  &&  FFM \cite{mccrae_introduction_1992}
                                                 &   GS \cite{bartneck_godspeed_2009} \\
        $\circ$ & added after pilot study               &&  TIPI \cite{gosling_TIPI_2003}
                                                 &   VA \cite{voelkel_developing_2020}\\
          &      &&   BFI-10 \cite{rammstedt_bfi-10_2007} &       \\        \\
     \end{tabular}
    \caption{List of attributes compiled from literature and pilot study. Part 4/4, showing a subset of the original 260 attributes that was not used by participants in the labeling task.}
    \label{stab:260_attributes_4}
\end{table}

\begin{table}[]
    \centering
    \begin{minipage}[t]{0.45\textwidth}
    \begin{tabular}{ll|cc}
        \textbf{source(s)}	&	\textbf{adjective} & \multicolumn{2}{c}{\textbf{in STEP}}	\\
        \hline \hline							
        GS, VA	&	friendly	&	$\square$	&	$\triangledown$	\\
        GS, VA'	&	mechanical	&	$\square$	&	$\triangledown$	\\
        AD, TIPI, VA'	&	simple	&	$\square$	&	$\triangledown$	\\
    	GS' &	robotic	&	$\square$	&	$\triangledown$	\\
    	&	creepy	&	$\square$	&	$\triangledown$	\\
    	&	weird	&	$\square$	&	$\triangledown$	\\
        \hline
        VA	&	playful	&	$\square$	&		\\
        GS, VA	&	intelligent	&	$\square$	&		\\
        VA	&	helpful	&	$\square$	&		\\
        GS	&	humanlike	&	$\square$	&		\\
        TIPI	&	complex	&	$\square$	&		\\
        GS'	&	humanoid	&	$\square$	&		\\
        GS'	&	scary	&	$\square$	&		\\
        $\circ$	&	animallike	&	$\square$	&		\\
        $\circ$	&	cute	&	$\square$	&		\\
        	&	strange	&	$\square$	&		\\
        	&	futuristic	&	$\square$	&		\\
        	&	functional	&	$\square$	&		\\
         \\
         \\
         \\
         \\
    \end{tabular}
    \end{minipage}
    \hfill
    \begin{minipage}[t]{0.43\textwidth}
    \begin{tabular}{ll|cc}
        \textbf{source(s)} & \textbf{adjective} & \multicolumn{2}{c}{\textbf{in STEP}} \\
        \hline \hline
    
        TIPI, BFI-10	&	reserved	&		&	$\triangledown$	\\
        IASR-B5, FFM, VA	&	assertive	&		&	$\triangledown$	\\
        GS	&	unpleasant	&		&	$\triangledown$	\\
        PCPS*, HRG	&	unemotional	&		&	$\triangledown$	\\
        GS	&	artificial	&		&	$\triangledown$	\\
        VA	&	synthetic	&		&	$\triangledown$	\\
        AD	&	clear	&		&	$\triangledown$	\\
        GS, VA	&	fake	&		&	$\triangledown$	\\
        TIPI*	&	unenthusiastic	&		&	$\triangledown$	\\
        GS*	&	unnatural	&		&	$\triangledown$	\\
        AD*	&	unclear	&		&	$\triangledown$	\\
        AD'	&	boring	&		&	$\triangledown$	\\
        $\circ$	&	masculine	&		&	$\triangledown$	\\
        $\circ$	&	male	&		&	$\triangledown$	\\
        $\circ$	&	young	&		&	$\triangledown$	\\
        $\circ$	&	feminine	&		&	$\triangledown$	\\
        $\circ$	&	female	&		&	$\triangledown$	\\
        	&	echo	&		&	$\triangledown$	\\
        	&	accent	&		&	$\triangledown$	\\
        	&	distorted	&		&	$\triangledown$	\\
        	&	fast	&		&	$\triangledown$	\\
        	&	monotone	&		&	$\triangledown$	\\
    \end{tabular}
    \end{minipage}
    \begin{tabular}{llp{2cm}lll}    \\
         \multicolumn{2}{l}{\textbf{markers:}} &&  \textbf{sources: }  &                                   \\
         \\
         $\square$ & in STEP-images        &&  IASR-B5 \cite{trapnell_extension_1990}
                                                & HRG \cite{hofstee_integration_1992}
                                                 &   AD \cite{Hassenzahl_attrakdiff:_2003}\\
         $\triangledown$ & in STEP-voices  &&  FFM \cite{mccrae_introduction_1992}
                                                & PCPS \cite{deRaad_pan-cultural_2014}
                                                 &   AD-BG \cite{hassenzahl_beauty_2004}\\
                         &                &&  TIPI \cite{gosling_TIPI_2003}
                                                 &   GS \cite{bartneck_godspeed_2009} &\\
         *       & antonym in source     &&   BFI-10 \cite{rammstedt_bfi-10_2007} 
                                                &   VA \cite{voelkel_developing_2020} &\\        
         '       & synonym in source     &&   & \\
         $\circ$ & added after pilot study  &&   &\\
         \\
     \end{tabular}
    
    \caption{The 40 labels selected for rating the robots' images and the voices. Whenever possible, references that confirm them were added for those that were not in the original list of 260 attributes.}
    \label{stab:40_dimensions}
\end{table}

\begin{table}
    \centering
    \begin{tabular}{|c|rl|}
        \hline
        \multirow{7}{8em}{
          \centering
          \includegraphics[height=6.52em]{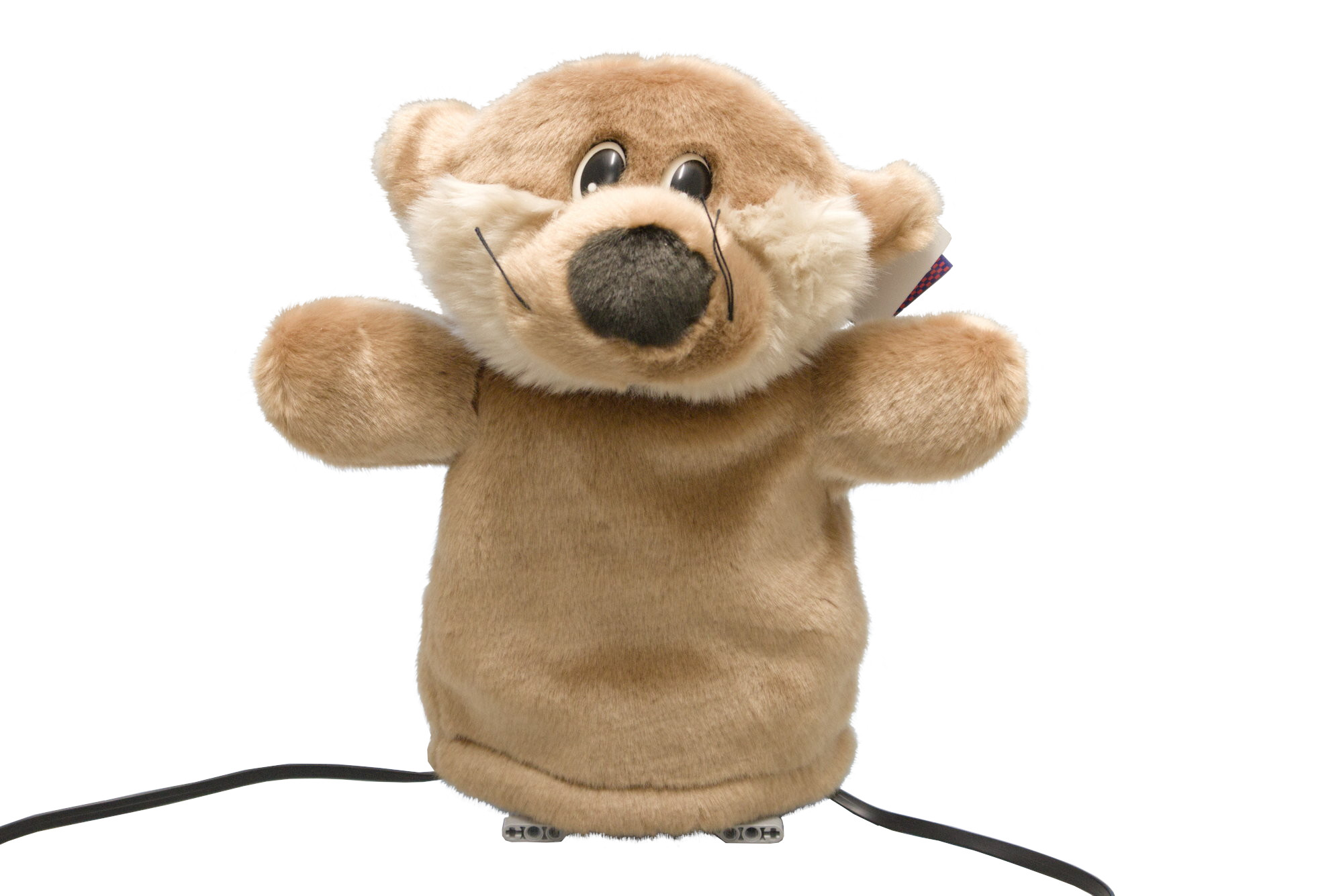}
        }   &&\\ &&\\ & \textbf{Name:} & B\"arbot \\ & \textbf{Creator:} & University of Augsburg \\ & \textbf{Image Credits:} & photo taken and edited by authors \\ &&\\ &&\\
        \hline
        \multirow{7}{8em}{
          \centering
          \includegraphics[height=9.0em]{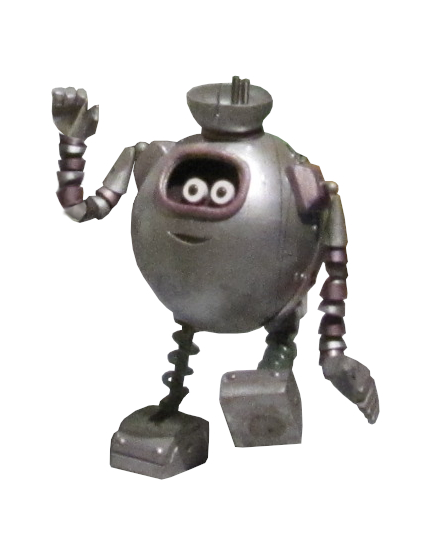}
        }   &&\\ &&\\ &\textbf{Name:} & Schlupp \\ & \textbf{Creator: } &Augsburger Puppenkiste \\ & \textbf{Image Credits:} & photo taken and edited by authors \\ &&\\ &&\\
        \hline
        \multirow{9}{8em}{
          \centering
          \includegraphics[height=11.0em]{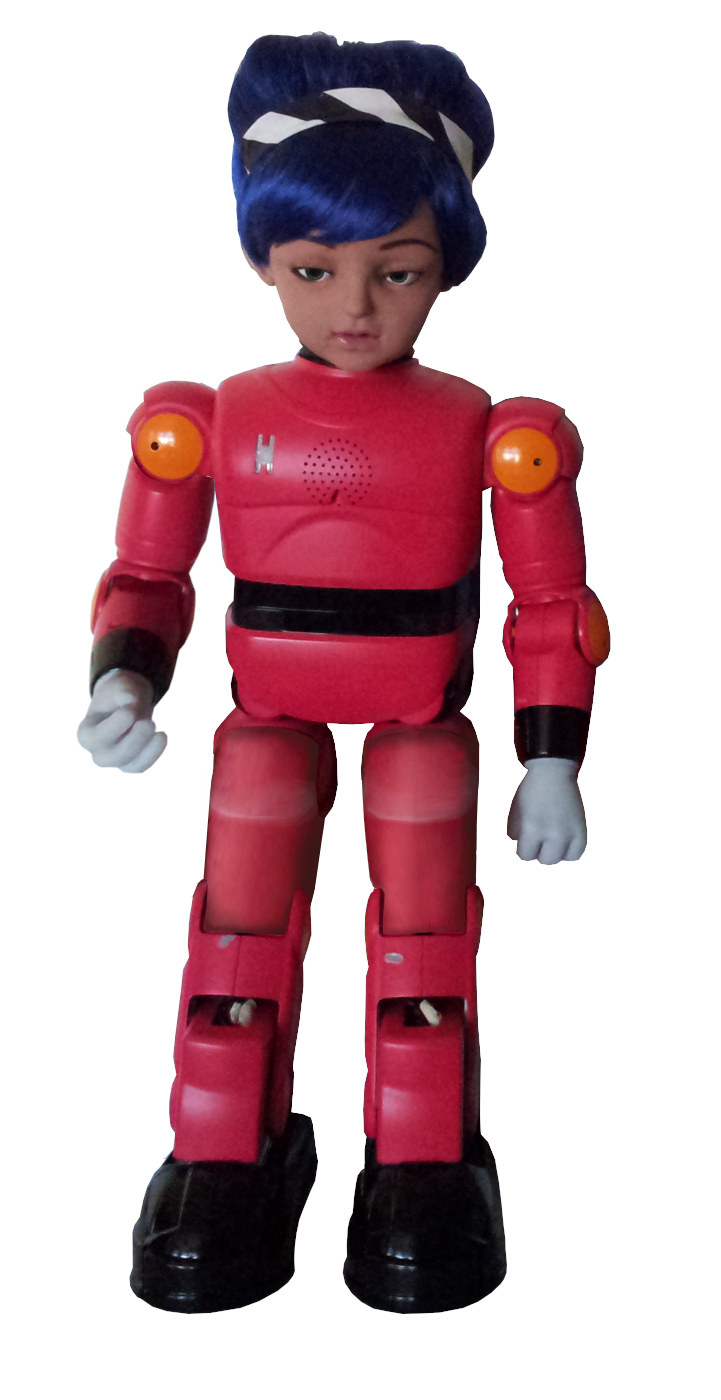}
        } &&\\ &&\\ &&\\  & \textbf{Name:} & RoboKind R50 Alice \\ & \textbf{Creator:} & Hanson Robotics \\ & \textbf{Image Credits:} & photo taken and edited by authors \\ &&\\ &&\\ &&\\ 
        \hline
        \multirow{7}{8.5em}{
          \centering
          \includegraphics[height=6.52em]{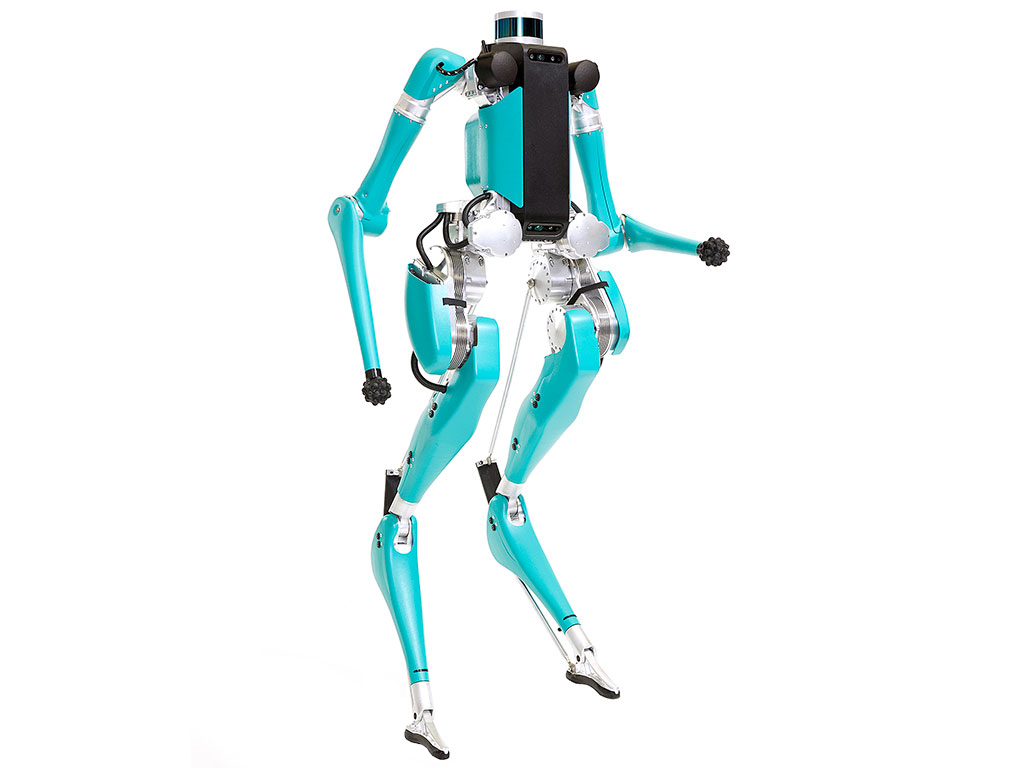}
        }   &&\\ &&\\ & \textbf{Name:} & Digit \\ & \textbf{Creator:} & Agility Robotics \\ & \textbf{Image Credits:} & Agility Robotics (written approval) \\ &&\\ &&\\
        \hline
        \multirow{9}{8em}{
          \centering
          \includegraphics[width=6em]{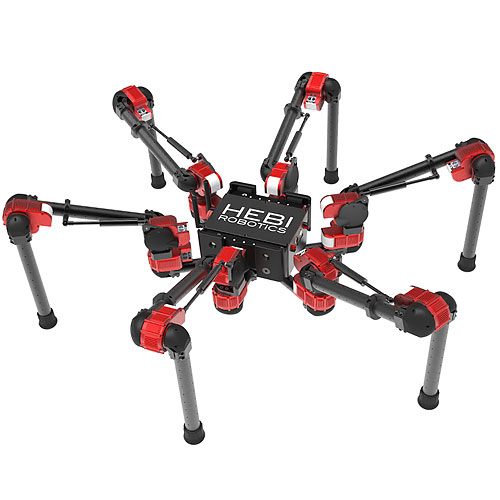}
        } &&\\ &&\\ &&\\  & \textbf{Name:} & Daisy \\ & \textbf{Creator:} & HEBI robotics \\ & \textbf{Image Credits:} & HEBI robotics (written approval) \\ &&\\ &&\\ &&\\ 
        \hline
        \multirow{7}{8.5em}{
          \centering
          \includegraphics[height=6.75em]{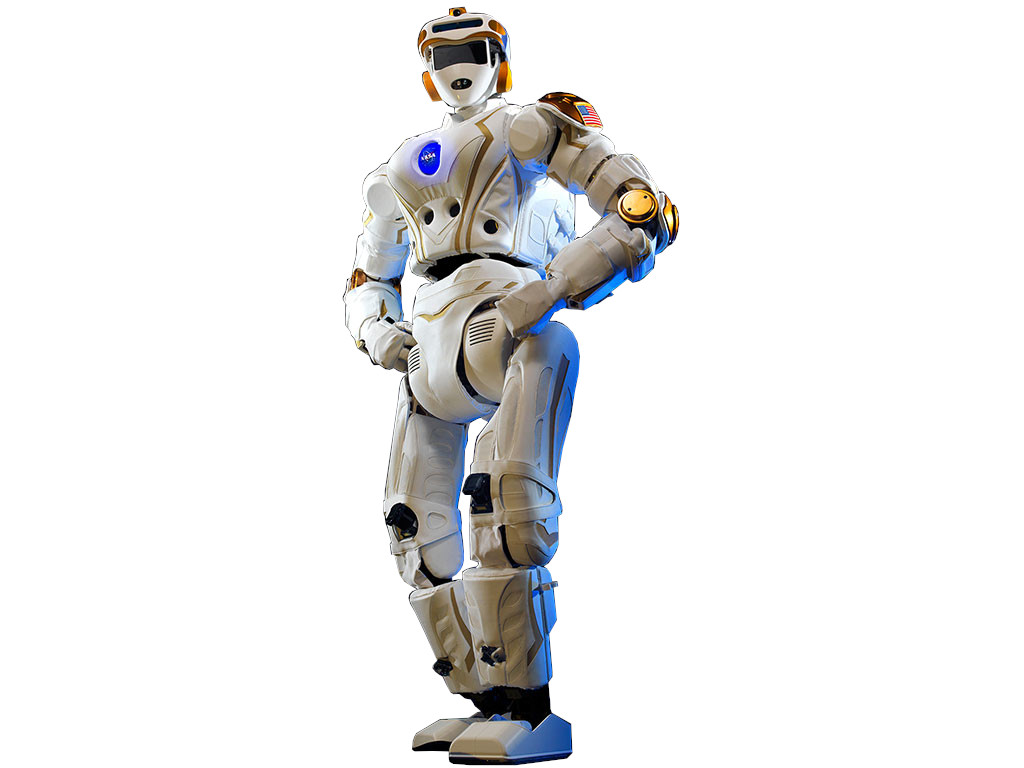}
        }   &&\\ &&\\ & \textbf{Name:} & Valkyrie \\ & \textbf{Creator:} & NASA\\ & \textbf{Image Credits:} & NASA (not subject to copyright for non-commercial use) \\ &&\\ &&\\
        \hline

        \multirow{7}{8.5em}{
          \centering
          \includegraphics[height=6.75em]{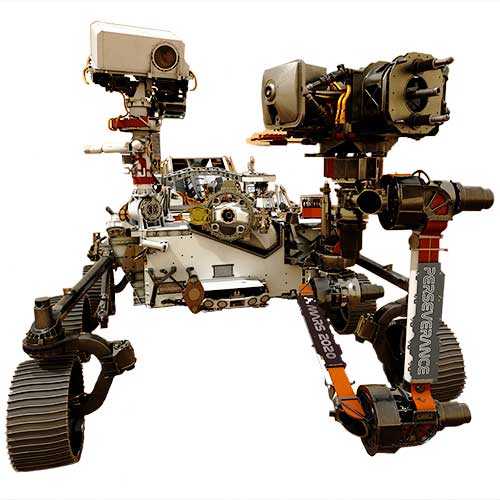}
        }   &&\\ &&\\ & \textbf{Name:} & Perseverance \\ & \textbf{Creator:} & NASA\\ & \textbf{Image Credits:} & NASA (not subject to copyright for non-commercial use) \\ &&\\ &&\\
        \hline

    \end{tabular}
    \caption{Copyright statement of robot images used in the figures of the manuscript (1/2).}
    \label{stab:robot_selection1}
\end{table}

\begin{table}
    \centering
    \begin{tabular}{|c|rl|}
    \hline
    \multirow{7}{8.5em}{
          \centering
          \includegraphics[height=6.75em]{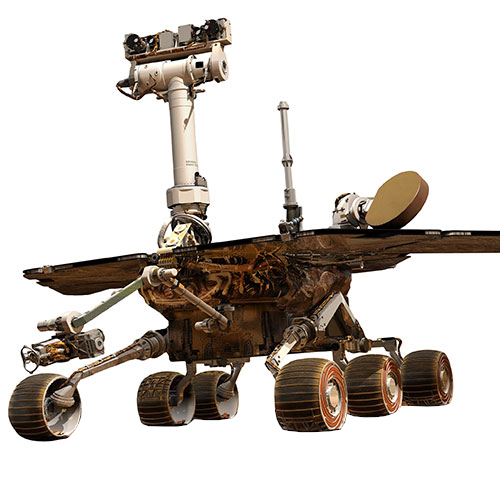}
        }   &&\\ &&\\ & \textbf{Name:} & Spirit \& Opportunity \\ \& \textbf{Creator:} & NASA\\ & \textbf{Image Credits:} & NASA (not subject to copyright for non-commercial use) \\ &&\\ &&\\
        \hline

        \multirow{7}{8.5em}{
          \centering
          \includegraphics[height=6.75em]{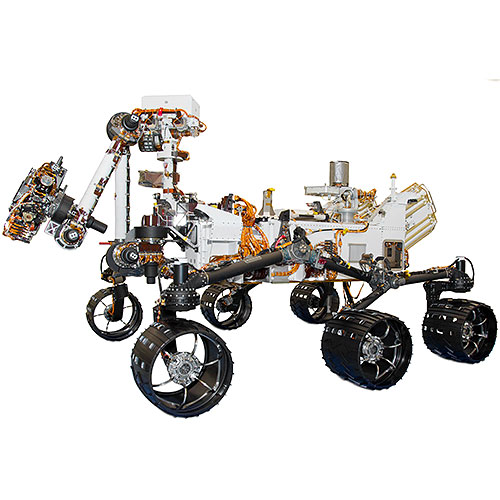}
        }   &&\\ &&\\ & \textbf{Name:} & Curiosity \\ & \textbf{Creator:} & NASA\\ & \textbf{Image Credits:} & NASA (not subject to copyright for non-commercial use) \\ &&\\ &&\\
        \hline

        \multirow{7}{8.5em}{
          \centering
          \includegraphics[height=6.75em]{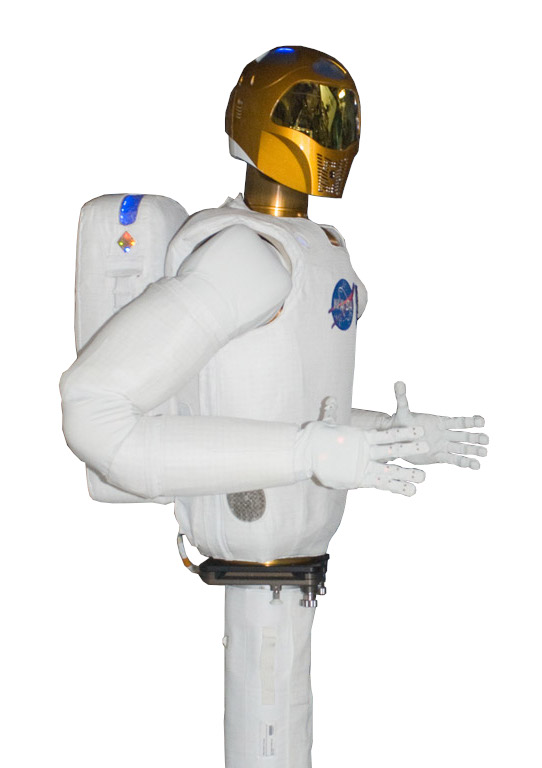}
        }   &&\\ &&\\ & \textbf{Name:} & Robonaut 2 \\ & \textbf{Creator:} & NASA\\ & \textbf{Image Credits:} & NASA (not subject to copyright for non-commercial use) \\ &&\\ &&\\
        \hline

    \end{tabular}
    \caption{Copyright statement of robot images used in the figures of the manuscript (2/2).}
    \label{stab:robot_selection2}
\end{table}

\end{document}